\documentclass[a4paper,fleqn]{cas-dc}
\usepackage{graphicx}
\usepackage[
    type={CC},
    modifier={by-nc-nd},
    version={4.0},
]{doclicense}
\usepackage{algpseudocode}
\usepackage{algorithm}
\usepackage{multicol}
\usepackage{longtable}
\usepackage{xcolor}
\usepackage{amsmath}
\usepackage{amssymb}
\usepackage{epstopdf}
\usepackage{subcaption}

\usepackage[numbers]{natbib}

\def\tsc#1{\csdef{#1}{\textsc{\lowercase{#1}}\xspace}}
\tsc{WGM}
\tsc{QE}
\tsc{EP}
\tsc{PMS}
\tsc{BEC}
\tsc{DE}

\begin{document}
\let\WriteBookmarks\relax
\def\floatpagepagefraction{1}
\def\textpagefraction{.001}

\shorttitle{Lightweight Mitigation Solution for DAO Insider Attack} 
\shortauthors{Verma et al.}

\title[mode=title]{Li-MSD: A Lightweight Mitigation Solution for DAO Insider Attack in RPL-based IoT}                     
\tnotemark[1]

\tnotetext[1]{https://doi.org/10.1016/j.future.2024.05.032}

\tnotetext[1]{A. Verma is Assistant Professor at Babasaheb Bhimrao Ambedkar University, Lucknow. S. K. Verma, and A. C. Pandey are with the Computer Science \& Engineering Discipline, PDPM Indian Institute of Information Technology, Design and Manufacturing, Jabalpur, Madhya Pradesh, India. G. Sharma is with the Department
of Computer Science \& Engineering, Malaviya National Institute of Technology Jaipur, Rajasthan, India. He is also Assistant Professor at Manipal University Jaipur.  J. Grover is with the Department
of Computer Science \& Engineering, Malaviya National Institute of Technology Jaipur, Rajasthan }

%

\affiliation[1]{organization={Computer Science \& Engineering Discipline, PDPM Indian Institute of Information Technology, Design and Manufacturing},
    addressline={Jabalpur}, 
    city={Madhya Pradesh},
    postcode={482005}, 
    country={India}}

\affiliation[2]{organization={Department of Information Technology, Babasaheb Bhimrao Ambedkar University},
	addressline={Lucknow}, 
	city={Uttar Pradesh},
	postcode={226025}, 
	country={India}}           

\affiliation[3]{organization={Malaviya National Institute of Technology},
            city={Jaipur},
            addressline={JLN Marg}, 
         citysep={}, 
            postcode={302017}, 
            state={Rajasthan},
            country={India}}
            
\affiliation[4]{organization={Manipal University Jaipur},
	city={Jaipur},
	addressline={Dehmi Kalan}, 
	citysep={}, 
	postcode={303007}, 
	state={Rajasthan},
	country={India}}

\author[1,2]{Abhishek Verma}[orcid=0000-0001-6687-4809]

\cormark[1]
\cortext[1]{Corresponding author}

\fnmark[1]

\ead{abhiverma866@gmail.com}


\credit{Conceptualization, Methodology, Validation, Investigation, Writing - Original Draft, Software, Supervision}

\author[1]{Sachin Kumar Verma}
\fnmark[2]
\ead{20mcs013@iiitdmj.ac.in}

\credit{Conceptualization, Resources, Validation, Visualization, Writing - Original Draft}

\author[1]{Avinash Chandra Pandey}[orcid=0000-0002-7831-4726]
\fnmark[3]
\ead{avish.p@iiitdmj.ac.in}

\credit{Formal analysis, Validation, Supervision}

\author[3]{Jyoti Grover}[orcid=0000-0001-9717-0441]
\fnmark[4]
\ead{jgrover.cse@mnit.ac.in}

\credit{Formal analysis, Validation, Data curation}

\author[3,4]{Girish Sharma}[orcid=0000-0001-6508-9438]
\fnmark[5]
\ead{2020rcp9012@mnit.ac.in}

\credit{Formal analysis, Validation, Data curation, Visualization}



\begin{abstract}
Many IoT applications run on a wireless infrastructure supported by resource-constrained nodes which is popularly known as Low-Power and Lossy Networks (LLNs). Currently, LLNs play a vital role in digital transformation of industries. The resource limitations of LLNs restrict the usage of traditional routing protocols and therefore require an energy-efficient routing solution. IETF's Routing Protocol for Low-power Lossy Networks (RPL, pronounced ``ripple") is one of the most popular energy-efficient protocols for LLNs, specified in RFC 6550. In RPL, Destination Advertisement Object (DAO) control message is transmitted by a child node to pass on its reachability information to its immediate parent or root node. An attacker may exploit the insecure DAO sending mechanism of RPL to perform ``DAO insider attack" by transmitting DAO multiple times. This paper shows that an aggressive DAO insider attacker can drastically degrade network performance. We propose a \underline{Li}ghtweight \underline{M}itigation \underline{S}olution for \underline{D}AO insider attack, which is termed as ``Li-MSD". Li-MSD uses a blacklisting strategy to mitigate the attack and restore RPL performance, significantly. By using simulations, it is shown that Li-MSD outperforms the existing solution in the literature.
\end{abstract}
\doclicenseThis




\maketitle

\section{Introduction}\label{Introduction}

\textcolor{black}{Routing Protocol for Low-Power and Lossy Networks (RPL) is currently a “Proposed Standard” and is still in its development stage \cite{rfc6550}. At present, there are many discovered and undiscovered (zero-day) vulnerabilities associated with RPL that an attacker may exploit to compromise protocol's performance. In the literature, many authors have argued that traditional cryptography-based solutions are not suitable for LLNs due to limited resources of nodes \cite{tsao2015security, sharma2023performance}. Traditional cryptography-based solutions occupy the most memory and take many CPU cycles, and consequently affect the overall performance of a resource-constrained nodes. Another reason for unsuitability of cryptography-based solutions in Low-Power and Lossy Networks (LLNs) is that such algorithms rely on the secure distribution of keys which is not an easy task when LLNs resource constraints are concerned \cite{el2016performance,ilia2013cryptographic, sharma2023qsec}. 
In a worst-case scenario, even if a single legitimate node is compromised, an attacker may gain access to a large pool of pre-loaded keys. It implies that all network nodes may get compromised once pre-loaded keys are revealed to attacker.The challenges related to secure key establishment, storage, distribution, revocation, and replacement in LLNs make traditional cryptography-based security solutions unsuitable for LLNs. In recent years, RPL has gained a lot of attention, and many new vulnerabilities have been identified, which has led to the significant increase of attack vector size \cite{bang2022assessment, mayzaud2017distributed}. One of such attacks against RPL is known as DAO insider attack \cite{ghaleb2018addressing, wadhaj2020mitigation}. The main motivation of attacker behind such DAO based attacks is to negatively affect the network capability to deliver packets from client nodes to server nodes, increase delay in packet delivery, and increase power consumption to reduce network lifetime. These parameters are very crucial for several IoT applications like smart agriculture, smart healthcare, and other mission critical applications. In the view of above mentioned security issue, the significant contributions of this paper are summarized below:}

\begin{enumerate}
	\item To address DAO insider attacks, an effective solution named ``Li-MSD" is proposed to mitigate the effects of attack in RPL-based IoT.
	\item Impact of DAO insider attacks on both static and mobile network scenarios is studied.
	\item Performance of Li-MSD is compared with the existing solution, i.e., SecRPL.
	\item \textcolor{black}{It is shown that Li-MSD achieves better false positive rate as compared to SecRPL.} 
	\item Using memory overhead as a parameter, it is shown that Li-MSD is a lightweight security solution.
\end{enumerate}

\textcolor{black}{In RPL, Destination Advertisement Object (DAO) messages are used to build downward routes. Downward routes enable the routing of data packets from a parent to a child node. It is the responsibility of child nodes to identify and let know the reverse route information to the DODAG root. To achieve this, RPL operates in two different modes, i.e., storing (table-driven) and non-storing (source routing) \cite{mishra2022hybrid}. In storing mode, the child node unicasts the DAO message to its parent, which then stores the received message. Then, the parent creates a new DAO message containing aggregate reachability information and sends it to the parent. Every intermediate parent performs this process until the DAO reaches the DODAG root. In non-storing mode, the intermediate parents don't store the DAO received from child nodes but only insert their own addresses to the reverse route stack. As per RFC 6560, it is necessary to create and maintain downward routes if Mode of Operation (MOP) is set to a non-zero value, i.e., 1 (Non-Storing), 2 (Storing Mode with no multicast support), or 3 (Storing Mode with no multicast support). MOP value of 0 resembles that DAO messages are disabled and only upward routes are supported by \textit{RPLInstance}. RFC 6550 does not specify how often the nodes must transmit these DAO messages. This is why different RPL implementations (i.e., ContikiRPL, OpenWSN, RIOT, Contiki-NG, OMNeT++, NetSim) choose different mechanisms to control DAO transmission rate. \textcolor{black}{In this paper, ContikiRPL is considered as it is the most widely used RPL implementation \cite{kim2017challenging}}. In ContikiRPL, the transmission of DAO messages is controlled using Trickle Timer. In RPL, DAO messages are unicast by the child node to the parent node basically on three occasions: when a node receives a DIO message from a parent node, when a node changes its preferred parent when a node detects some routing error. However, an attacker can take advantage of the non-secured operation of RPL to disrupt the normal network's performance by simply transmitting malformed or eavesdropped DAO message very frequently to its parent. The best-case scenario for an attacker will be to launch the attack from the edge of the network as this will increase the number of DAO messages getting transmitted. This type of attack is known as DAO insider attack, and it can be launched using insider as well as outsider attack strategy \cite{ghaleb2018addressing}. There has been advancement in DAO based attacks like Dropped DAO attack \cite{sheibani2022lightweight, goel2023cra, goel2023lightweight}, Network partitioning attack \cite{sahay2022mitigating}, and DAO induction attack \cite{baghani2021dao}, however the scope of this paper is limited to DAO insider attack only.} 
	
\textcolor{black}{Ghaleb et al. \cite{ghaleb2018addressing} proposed a defense solution named SecRPL to address DAO insider attack, however the solution has many limitations including frequent unblocking of attacker node, lack of early mitigation logic, blocking of victim nodes, and only static networks based evaluation. To address DAO insider attacks, we propose a lightweight security solution named ``Li-MSD" which addresses all the research gaps induced by SecRPL. Li-MSD is a threshold based distributed attack detection technique which uses early blacklisting strategy to quickly mitigate the attack one they are detected, and consequently saves node's computational resources. Li-MSD is implemented on Contiki operating system and evaluated on Cooja simulator for both static and mobile network scenarios. This paper is an extended version of our previous work \cite{Verm2206:Addressing}.} 


\textcolor{black}{A concise statement of research problem considered in this paper is mentioned below.} 


\subsection*{\textcolor{black}{Research Problem Statement}}
\textcolor{black}{\textit{The increasing prevalence of DAO insider attacks on RPL poses a significant threat to the performance and security of LLNs. The DAO insider attack exploits vulnerabilities in RPL by replaying eavesdropped or malformed messages, thereby jeopardizing the overall performance of LLNs. Despite the identification of this threat in previous research \cite{ghaleb2018addressing}, the lack of inbuilt lightweight security mechanisms in RPL remains a critical issue. This research aims to address the gap in current security measures for RPL and develop effective lightweight security mechanisms to mitigate the impact of DAO insider attacks on LLNs.}}


The rest of this article is organized as follows. Section \ref{Background} discusses the basics of RPL protocol and illustrates DAO insider attack. Section \ref{Related Work} overviews the related works. Further, Section \ref{Proposed Lightweight Mitigation Solution} presents the proposed lightweight mitigation solution Li-MSD. Section \ref{Performance Evaluation} emphasizes on simulation metrics and performance analysis. Lastly, the paper is concluded in Section \ref{Conclusion and Future Scope}.

\section{Background}\label{Background}
\subsection{Overview of RPL Protocol}

\textcolor{black}{RPL is a proactive routing protocol that uses source routing and distance vectors. RPL is designated as a ``Proposed benchmark” in RFC 6550 \cite{rfc6550}. As RPL takes less energy to set up and maintain the network topology, it is considered an energy-efficient protocol \cite{gaddour2012rpl}. For routing, it employs the distance vector protocol. RPL uses the top layer of IEEE $ 802.15.4 $ MAC. RPL creates a topology called DODAG (Destination Oriented Directed Acyclic Graph) on LLN devices. DODAG is a tree-like structure with no loops in which the root node is considered the destination for all nodes. At any given time, the network may be running many DODAGs, all of which are identified as an \textit{RPLInstance}. \textit{RPLInstance}  is recognized by \textit{RPLInstanceID}, which is a unique IPv6 address.
A rank value is allocated to each LLN node, which is a 16-bit integer that shows the node position in relation to the DODAG root node. 
The idea of rank is employed for the following reasons.} 

\begin{itemize}
	\item To repair the damaged links.
	\item To distinguish between parents and siblings.
	\item To identify and fix routing loops.
	\item To establish a connection between child and parent.
\end{itemize}

DODAG Information Solicitation (DIS), DODAG Information Object (DIO), Destination Advertisement Object (DAO), and Destination Advertisement Object Acknowledgment (DAO-ACK) are the four types of control messages supported by RPL. 
For rank calculation, RF defines an Objective Function (OF) \cite{lamaazi2020comprehensive}. OF is used to find the best parent with the shortest path to the DODAG root node. There are a variety of objective functions available, including (1) MRHOF\cite{gnawali2012minimum}, (2) ETX OF\cite{gnawali2010etx}, (3) OF0\cite{thubert2012objective}, and many others. A node utilizes the DIS message to locate and join a DODAG that already exists. In RPL, the DIO message is used for broadcasting. To create a new DODAG, the root node sends a DIO message. The DIO message contains the relevant information, which another node can utilize to discover the DODAG. The DIO message helps create upward routes from child nodes to the DODAG root, while the DAO message is used to create downward routes. 
Every node except the root node unicasts a DAO message to advertise their addresses and prefixes to their parent. After that, a DAO Recipient sends a DAO-ACK unicast message to nodes from which they received the DAO message. The concept of  ``Trickle Timer” is employed to reduce the control message transmissions by sensor nodes \cite{levis2011trickle}. The interval of the trickle timer may be increased or decreased in case of any inconsistency or static network.

\subsection{DAO Insider Attack in RPL}
\textcolor{black}{To get better insight into the DAO insider attack, we have explained the attack behavior w.r.t. storing  and non-storing mode of RPL. The main differences between storing mode and non-storing mode of RPL are listed below.} 

\subsection*{\textcolor{black}{Storing Mode}}
\begin{itemize}
	\item \textcolor{black}{Some nodes (storing nodes) store complete network topology information.}
	\item \textcolor{black}{Suitable for devices with more resources.}
	\item \textcolor{black}{Storing nodes make routing decisions based on stored information.}
\end{itemize} 
\subsection*{\textcolor{black}{Non-Storing Mode}}
\begin{itemize}

	\item \textcolor{black}{Each node stores only its parent node in the routing tree.}
	\item \textcolor{black}{Suitable for resource-constrained devices.}
	\item \textcolor{black}{Routing decisions are made dynamically based on local information.}
\end{itemize}
\textcolor{black}{The choice between storing and non-storing modes depends on the characteristics of the network and the devices involved.} 

\subsubsection{DAO Insider Attack in Storing Mode}
\begin{figure}[!htb]
	\centering
		\includegraphics[width=0.52\textwidth, trim={0cm .5cm 0cm  0.5cm}]{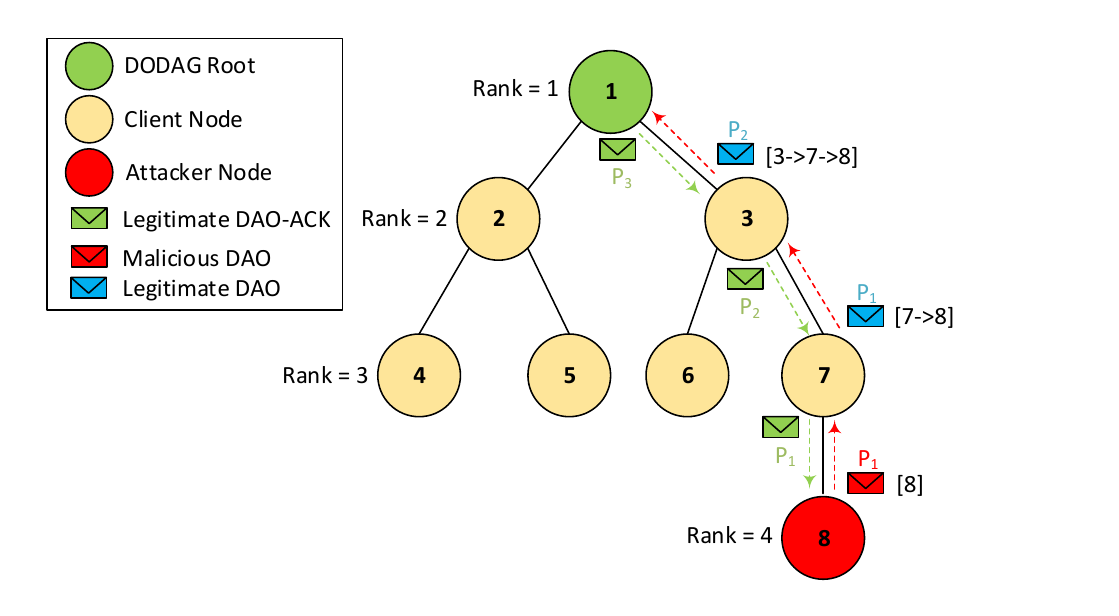}
		\caption{DAO insider attack in storing mode}
		\label{fig:DAO_S}
 \end{figure}

As shown in Fig. \ref{fig:DAO_S}, the attacker node with ID 8 is sending a malicious DAO message to its parent with ID 7. The parent node stores the received DAO message in storing mode and creates a new DAO message containing aggregate reachability information. All the new packets created by the intermediate parents (Node ID 7 and 3) are legitimate and transmitted upward in a unicast manner. Upon receiving a DAO from the child node, the parent node or DODAG root replies with a unicast DAO-ACK message (\textcolor{LimeGreen}{$P_{1}$,$P_{2}$,$P_{3}$}) which is forwarded by the child nodes in downward direction \cite{gaddour2012rpl}. It is to be noted that a single malicious DAO transmission (\textcolor{red}{$P_{1}$}) by the attacker node triggers the creation and transmission of legitimate DAO messages (\textcolor{Cerulean}{$P_{1},P_{2}$}) equal to the number of intermediate parents between attacker and DODAG root. The attacker can take advantage of this nature of RPL to disrupt the normal network operation and negatively impact network's performance by repeatedly transmitting DAO messages to its parent.    

 \begin{figure}[!htb]
		\includegraphics[width=0.52\textwidth, trim={0cm .5cm 0cm  1cm}]{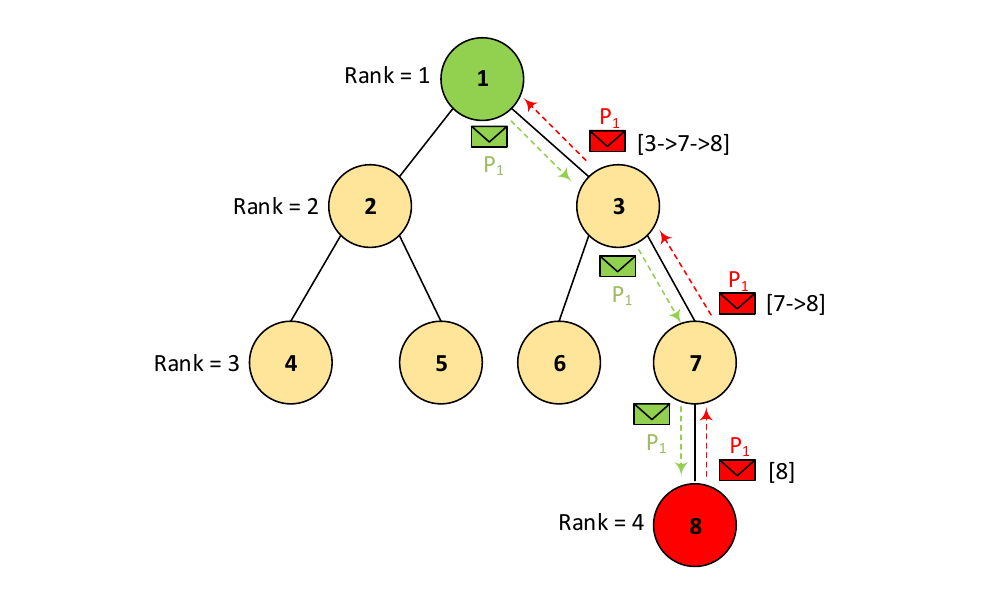}
		\caption{DAO insider attack in non-storing mode}
		\label{fig:DAO_NS}
\end{figure}

\subsubsection{DAO Insider Attack in Non-Storing Mode}
In non-storing mode, the intermediate parents do not store DAO messages received from child nodes. The parent node forwards the DAO message to its own parent after appending its own address to the reverse route stack. DAO insider attack is possible in this scenario also. As shown in Fig. \ref{fig:DAO_NS}, the attacker (Node ID 8) sends a malicious DAO message (\textcolor{red}{$P_{1}$}) to its parent (Node ID 7), which further forwards it to its own parent after appending its own address (let us say the Node ID represent the address) to the DAO message. Unlike storing mode, intermediate parents do not create any new DAO messages in this case. Upon receiving a DAO message from the child node, the DODAG root responds with a unicast DAO-ACK message (\textcolor{LimeGreen}{$P_{1}$}) destined to Node 8. Although that attacker is not able to induce the creation of a new DAO message at intermediate parents, it may still succeed in bringing down the network performance. The impact of a DAO insider attack depends on the rate of malicious DAO transmission by the attacker node. 

\textcolor{black}{LLN-based IoT applications, such as smart home and smart agriculture, involve the use of downward routing (i.e., packets needs to be routed from client nodes to server nodes), making the use of DAO packets necessary. An attacker may see this as an opportunity to launch a DAO insider attack on the network, negatively impacting its capability and performance in delivering data packets to the gateway router. This situation becomes more critical when applications cannot tolerate large delays in data packet delivery. Considering this, it can be concluded that addressing DAO insider attacks is crucial. The Li-MSD design and mitigation logic not only detect the attack but also enable quick mitigation after the first detection. More details about the working of Li-MSD are discussed in Section \ref{Proposed Lightweight Mitigation Solution}.}

\section{Related Work}\label{Related Work}
\textcolor{black}{In this section, we discuss some recent works related to the security of RPL. These solutions can be classified in to two broad categories i.e., solutions for DAO insider attack mitigation, and solutions with no mitigation of DAO insider attack. The latter classification includes solutions that lay the basis of defense solution proposed in this paper.} 

\subsection{Solutions for DAO Insider Attack Mitigation} 

Ghaleb et al. \cite{ghaleb2018addressing} addressed DAO insider attack by putting the restriction on the number of DAO messages forwarded per destination. The proposed solution SecRPL uses a threshold parameter and a DAO counter that monitors the number of DAO received from child nodes. When the threshold is reached for any child node, the parent blocks it and stops forwarding further DAO for a particular period of time. To ensure that no child remains permanently blocked due to the time factor, the DAO counter is reset (for all child nodes) when the parent sends a new DIO. The significant limitations of SecRPL are: (1) it resets the DAO counter between two consecutive DIO transmissions for all the child nodes, which may lead to unblocking of the attacker node frequently. In the worst-case scenario, the attacker may not be even blocked if SecRPL is deployed in mobile networks where DIO transmissions are frequent; (2) SecRPL lacks early mitigation of attack that can save a lot of node's resources that are consumed during unnecessary processing of DAO messages at the network layer; (3) the detection algorithm is not capable of distinguishing between a victim and attacker node; therefore, it may sometime block legitimate nodes also; (4) it is tested only for static networks. \textcolor{black}{Wadhaj et al. \cite{wadhaj2020mitigation} extended the work of Ghaleb et al. \cite{ghaleb2018addressing} and proposed SecRPL1, and SecRPL2. SecRPL1 works by limiting the number of forwarded DAOs per child, whereas in SecRPL2 the number of forwarded DAO's are restricted regardless of child who initiated the DAO. SecRPL1 blocks only identified malicious child nodes. In contrast. SecRPL2 blocks all the child nodes when attack is detected. Both SecRPL1 and SecRPL2 use static threshold and evaluated for static networks only. SecRPL1 performs better than SecRPL2.}      

\subsection{Solutions with no mitigation of DAO insider attack}

Pu et al. \cite{pu2022lightweight} proposed a lightweight security solution named \textit{liteSAD} for addressing sybil attack in RPL. The authors used a physical unclonable function (PUF) to uniquely identify each node and store its information in an array. To reduce the memory overhead of storing the addresses as a separate entry, \textit{liteSAD} uses a Bloom filter instead of an array. The major limitations of \textit{liteSAD} is the overhead induced by the Bloom filter itself and the issue of collision when the number of nodes increases in the network. Verma et al. \cite{verma2020mitigation} proposed a solution named \textit{Secure-RPL} for defending RPL against DIS attacks. \textit{Secure-RPL} uses RPL parameters and puts thresholds on them to identify attackers and block them. \textit{Secure-RPL} is unable to identify sybil attacker. This limitation is shown to be solved by the RPL-MRC approach proposed by Medjek et al. \cite{medjek2021multicast}. RPL-MRC is based on two complementary mitigation mechanisms: Response Delay and Timer Readjustment. RPL-MRC improves the RPL performance in terms of control packet overhead and power consumption. In \cite{sahay2022mitigating} an approach called Enhanced RPL (ERPL) is proposed to mitigate the worst parent attack. ERPL modifies the existing RPL process that populates the candidate neighbor set using rank value. Rank is used to develop an optimal parent set and ensure that the nodes choose the parent from this optimal set only. Bang et al. \cite{bang2022embof} proposed an Echelon Metric Based Objective Function (EMBOF–RPL) for identifying rank attacks. An Echelon value is injected into the DIO messages' DAG metric container and then transmitted to child nodes. When the sender multicast the DIO packet, the receiver first checks the legitimacy of the received rank and then calculates its own rank. Using simulations, it is shown that EMBOF-RPL outperforms ContikiRPL. Kaliyar et al. \cite{kaliyar2020lidl} addressed sybil and wormhole attacks using the concept of Highest Rank Common Ancestor (HRCA). The proposed approach named LiDL identifies the ongoing attack and locates the attacker's position. Sharma et al. \cite{sharma2022mitigation} proposed a two-step verification model based on distributed timer mechanism to address blackhole attacks. The proposed approach initially marks the parent as a potential attacker and then performs blackhole node verification with the help of neighboring nodes. Although the proposed solution achieves a true positive rate of 100\%, it is not tested for other metrics. Agiollo et al. \cite{agiollo2021detonar} proposed a attack detection approach named DETONAR which is based on the combination of signature and anomaly-based rules. The major limitation of this approach is that it uses a packet sniffer. Moreover, the packet sniffer needs to work in promiscuous mode to continuously listen to packet transmissions.

\clearpage
\onecolumn
\color{black}
\begin{longtable}{|p{1.5cm}|p{1.2cm}|p{1.6cm}|p{2.5cm}|p{2cm}|p{2cm}|p{0.5cm}|}
	\caption{\textcolor{black}{Comparison of existing solutions with our proposed solution}}
	\label{tab:comparision}\\
	\hline
	\textbf{Authors} & \multicolumn{1}{c|}{\textbf{\begin{tabular}[c]{@{}c@{}}Defense \\ Mechanism\end{tabular}}} & \multicolumn{1}{c|}{\textbf{\begin{tabular}[c]{@{}c@{}}Attack \\ Addressed\end{tabular}}} & \multicolumn{1}{p{1.5cm}|}{\textbf{\begin{tabular}[c]{@{}c@{}}Approach \\ Used\end{tabular}}} & \multicolumn{1}{c|}{\textbf{Advantages}} & \multicolumn{1}{c|}{\textbf{Limitations}} & \multicolumn{1}{c|}{\textbf{\begin{tabular}[c]{@{}c@{}}Mitigation\\of DAO \\ Insider Attack\end{tabular}}} \\ \hline
	\endhead
		Pu et al. \cite{pu2022lightweight} & \multicolumn{1}{c|}{liteSAD} & \multicolumn{1}{c|}{Sybil} & Physical Unclonable Function (PUF) with Bloom filter & Lightweight Solution, Incorporates PUF for additional physical security, Scalable & False positives, Dependency on PUF, Network overhead, Limited to specific IoT scenarios & No  \\ \hline
		Verma et al. \cite{verma2020mitigation} & \multicolumn{1}{c|}{Secure-RPL} & \multicolumn{1}{c|}{DIS Flooding} & Thresholding of RPL configuration parameters & Lightweight solution, Fast detection, Scalable & Implementation overhead, Static thresolding, Not effective in slow DIS attacks & No  \\ \hline
		Medjek et al. \cite{medjek2021multicast} & \multicolumn{1}{c|}{RPL-MRC} & \multicolumn{1}{c|}{DIS Flooding} &  Response delay with timer readjustment & Scalable, Low implementation overhead  & Negatively impact convergence time of the network & No \\ \hline
		Sahay et al. \cite{sahay2022mitigating} & \multicolumn{1}{c|}{ERPL}  & \multicolumn{1}{c|}{Worst Parent} & Strict rank optimality & Low communication overhead, scalable & Cannot detect other rank based attacks & No \\ \hline
		Bang et al. \cite{bang2022embof} & \multicolumn{1}{c|}{EMBOF-RPL} & \multicolumn{1}{c|}{Increased Rank} & Echelon Metric Based Objective Function & Isolation of attacker, limited scalability   & Vulnerable against coordinated attack, requires architectural changes in RPL & No \\ \hline
		Kaliyar et al. \cite{kaliyar2020lidl} & \multicolumn{1}{c|}{LiDL}  & \multicolumn{1}{c|}{Wormhole} & Highest Rank Common Ancestor  & Localization of attacker, lightweight and fast detection, scalable, low overhead & Evaluated for a small simulation time & No \\ \hline
		Sharma et al. \cite{sharma2022mitigation} & Distributed timer-based mechanism & \multicolumn{1}{c|}{Blackhole} & High detection accuracy, Scalable, Low Overhead &  & Performance is dependent on an appropriate value of the threshold & No \\ \hline
		Agiollo et al. \cite{agiollo2021detonar} & \multicolumn{1}{c|}{DETONAR} & Blackhole, Clone ID, Sinkhole,
		DIS Flood, Hello Flood, Local Repair, Increased Rank, Replay,  Selective Forwarding, Sybil, Version number, Wormhole, Worst Parent & Signature and Anomaly Detection & Multiple attack detection, Low false postive rate, No overhead on nodes  & Non-scalable, Requires additional sniffer nodes, High power consumption, Costly solution & No \\ \hline
		Ghaleb et al. \cite{ghaleb2018addressing} & \multicolumn{1}{c|}{SecRPL} & \multicolumn{1}{c|}{DAO Insider Attack} & Thresholding & Low implementation overhead, Scalable & Frequent unblocking of attacker node, Lack of early mitigation logic, Suffers from victim node problem, tested for only static networks & Yes  \\ \hline
		Wadhaj et al. \cite{wadhaj2020mitigation} & \multicolumn{1}{c|}{\begin{tabular}[c]{@{}c@{}}SecRPL1\\SecRPL2\end{tabular}} & \multicolumn{1}{c|}{DAO Insider Attack} & Thresholding & Low implementation overhead, Scalable & Frequent unblocking of attacker node, Lack of early mitigation logic, Suffers from victim node problem, tested for only static networks & Yes  \\ \hline
		Our Proposed Solution & \multicolumn{1}{c|}{Li-MSD} & \multicolumn{1}{c|}{DAO Insider Attack} & Thresholding with early blacklisting strategy & Solves victim node problem, improves network performance, detection with mitigation, lightweight solution, scalable  & Implementation overhead, static threshold  & Yes \\ \hline
\end{longtable}
\clearpage
\twocolumn
\color{black}

\textcolor{black}{Table \ref{tab:comparision} provides a detailed comparison of the some exiting defense solutions in the literature. Major limitations of RPL based defense solutions like liteSAD and DETONAR is the additional implementation overhead which add costs to the solution. DETONAR requires an additional sniffer node and its strategic placement for proper monitoring of network activities. Secure-RPL and Distributed Timer based mechanisms are dependent on configuration of appropriate threshold value. EMBOF-RPL requires architectural changes to RPL which may affect normal RPL operations and performance. Most of the solutions are only tested on static networks scenarios. The existing solution for DAO insider attack (i.e., SecRPL) has many shortcomings as mentioned earlier. In this work, we have focused on developing a solution named Li-MSD to address DAO insider attack that considers both static and mobile networks while removing shortcomings of SecRPL. } 

\section{Proposed Lightweight Mitigation Solution}\label{Proposed Lightweight Mitigation Solution}
\textcolor{black}{Li-MSD is proposed to mitigate the DAO insider attack in RPL-based IoT networks. The main objective of Li-MSD is to monitor the behavior of children nodes by analyzing number of DAO messages received from the child nodes. Li-MSD checks for the original source of the DAO message before deciding the attacker node. Once the attacker node is identified Li-MSD adds it to blacklist to avoid any further communication with attacker node.}

\begin{figure*}[!htb]
	\centering
	\centering
	\includegraphics[width=0.7\textwidth]{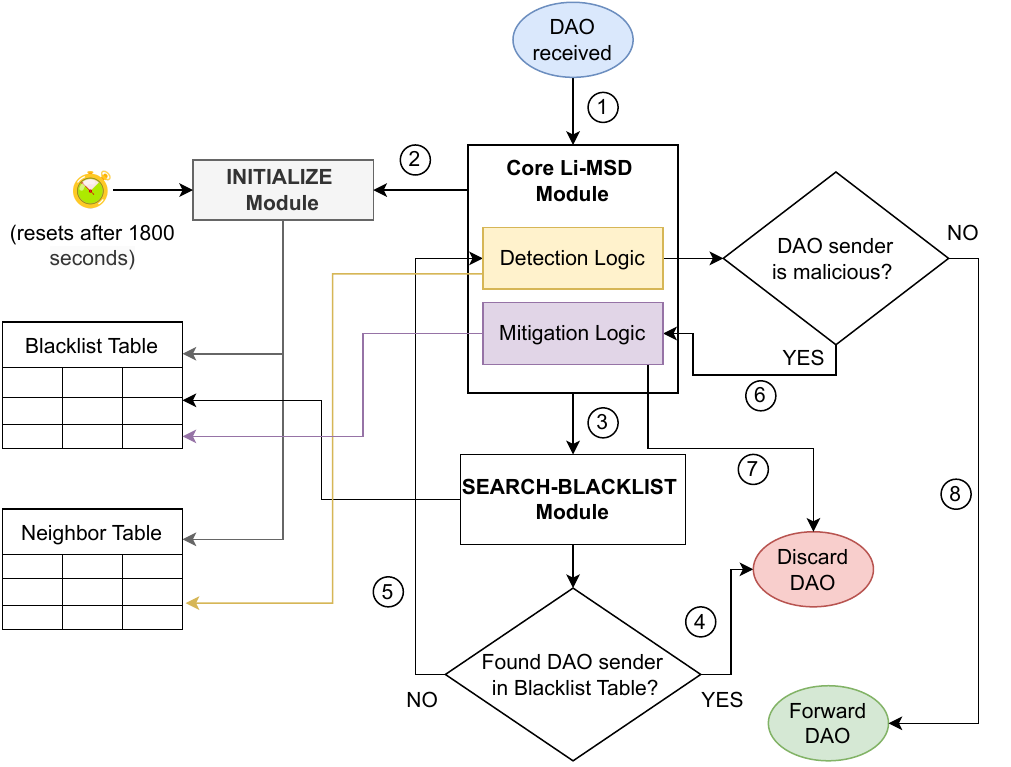}
	\caption{\textcolor{black}{Architecture of Li-MSD}}
	\label{fig:architecture}
\end{figure*}

\textcolor{black}{Li-MSD is based on distributed detection strategy, which involves every sensor node running its instance of defense algorithm. The node running Li-MSD maintains two tables, i.e., a neighbor table (\textit{$\mathcal{Q}$}) for storing information about child nodes and a blacklist table (\textit{$\mathcal{Z}$}) to store information about blacklisted child nodes (identified attacker nodes. A major benefit of using blacklist table is that it helps in saving node computational resources by preventing unnecessary processing of DAO messages received from blacklisted nodes. Li-MSD uses $\beta$, i.e., a threshold to limit the maximum number of DAO messages that can be forwarded for a particular child node. \textcolor{black}{The value of $\beta$ is chosen based on the analysis of multiple non-attack scenarios, and it is configured at the time of node deployment. A smaller value of $\beta$ reduces attack detection time but increases false positives, and vice-versa. Therefore $\beta$ must be set carefully depending on the size of target network.} Li-MSD method consists of three procedures: INITIALIZE, SEARCH-BLACKLIST, and Li-MSD. Pseudocodes of listed procedures are shown in Algorithms \ref{Mitigation procedure}, \ref{Initialization procedure}, \ref{Search blackilist procedure}. The architecture of Li-MSD is depicted in Fig. \ref{fig:architecture}. Flows (numbered from 1-8) are mentioned in the Fig. \ref{fig:architecture} for better understanding of sequence of events in Li-MSD. Table \ref{Symbols} presents different symbols (data structures, variables) and corresponding definitions used in the proposed solution. }

\begin{figure}[!htb]
	\centering
	\centering
	\includegraphics[width=0.5\textwidth]{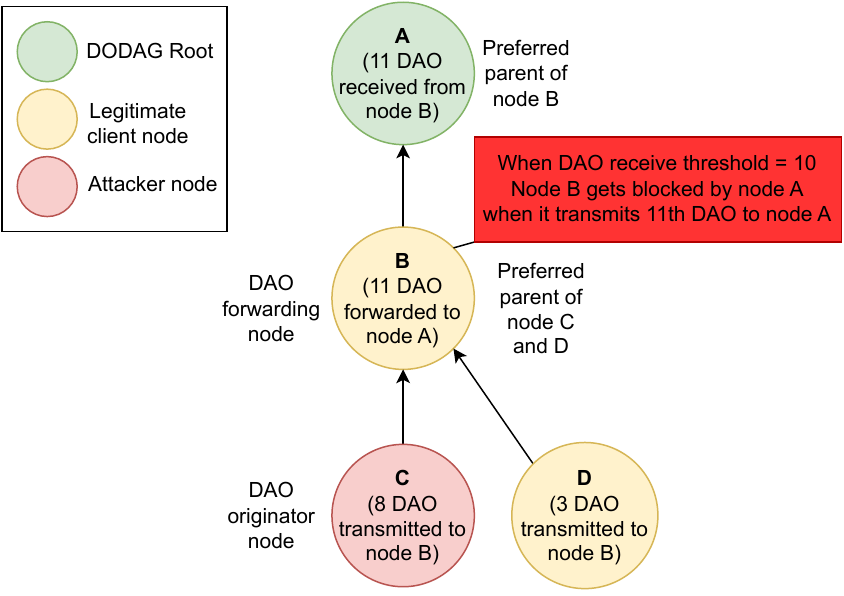}
	\caption{\textcolor{black}{An Illustration of Victim Node Problem (DAO receive threshold = 10)}}
	\label{fig:VNP}
\end{figure}

\begin{figure}[!htb]
	\centering
	\centering
	\includegraphics[width=0.5\textwidth, trim={2cm 2cm 2cm 2cm}]{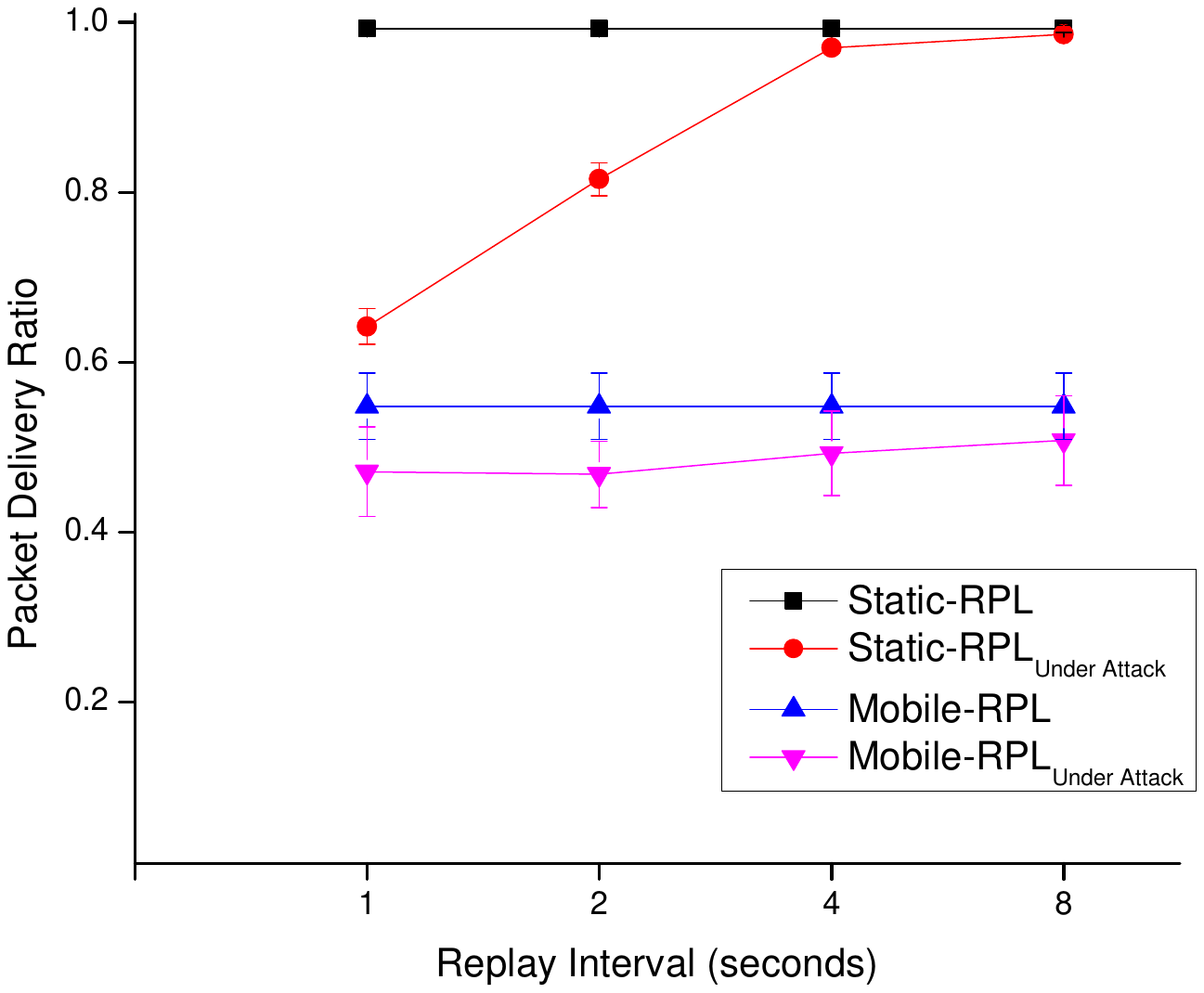}
	\caption{\textcolor{black}{Comparison of packet delivery ratio in non-attack and attack network scenarios}}
	\label{fig:u_ua_comp}
\end{figure}

The key idea is distinguishing between legitimate, attacker, and victim nodes. There are many cases when legitimate appears to be an attacker node, which is actually a victim node. This situation leads to the blocking of victim nodes, which increases the false positives of the defense solution. SecRPL \cite{ghaleb2018addressing} is one solution in literature that addressed DAO insider attack. SecRPL suffers from the problem of blocking victimized nodes. Li-MSD addresses important research gaps like reducing false positives, early attack mitigation, and mobility support. At first, we conducted multiple experiments with different network scenarios, including both non-attack and under-attack, to analyze the behavior of legitimate (normal) and attacker nodes in terms of the number of RPL control messages transmitted and received by them. As this study focuses on DAO insider attacks, special attention is given to analyzing the behavior of nodes in terms of the number of DAO messages being transmitted and received. With a detailed analysis, we conclude that each node in the RPL network receives and transmits a similar number of DAO messages in non-attack scenarios. \textcolor{black}{Whereas, in the case of an under-attack scenario, a legitimate node receives a comparatively large number of DAO messages from a malicious node. Fig. \ref{fig:u_ua_comp} shows the packet delivery ratio obtained in non-attack (Static-RPL and Mobile-RPL) and attack (Static-RPL\textsubscript{Under Attack} and Mobile-RPL\textsubscript{Under Attack}) scenarios. It can be observed that packet delivery ratio in non-attack static and mobile scenarios is always higher that under-attack static and mobile scenarios.} This analysis helps to develop a solution that can easily identify the attacker node among a set of neighbors. \textcolor{black}{Considering the difference between number of DAOs transmitted in under-attack and non-attack scenario a straightforward solution to mitigate this problem is to simply block the the DAO forwarding of malicious child nodes. But, there is a problem with such kind of solution which we call as ``The Victim Node Problem".} As RPL-based networks involve multi-hop transmission, a node might fall into the category of attacker node, although it is simply forwarding the packets received from the attacker node. Therefore such node becomes the victim of the attack and possibly gets blocked by its parent. \textcolor{black}{Fig. \ref{fig:VNP} illustrates the ``Victim Node Problem" where value of DAO receive threshold (maximum DAO a node can transmit in a given period of time) is set to 10. The illustration shows that Node B (a legitimate node) gets blocked by Node A even when it is not the originator of malicious DAOs because it has reached maximum DAO receive threshold. This situation occurred because total DAOs forwarded by Node B to Node A have crossed DAO receive threshold, however none of its child nodes (Node C and D) have crossed DAO receive threshold. In this case Node B is victim and consequently legitimate packets of Node D will also be blocked further.} We have solved the ``The Victim Node Problem" by simply using the variables of RPL protocol. In further explanation, we will show how ``The Victim Node Problem" is addressed by Li-MSD.

\textcolor{black}{An important point to note is that there can be non-attack situation where large of DAOs may be transmitted. In this case the network performance will be degraded similar to that of an under-attack scenario. However considering the fact that DAO is transmitted by child node to its preferred parent node upon receiving a DIO from that parent, upon preferred parent change, and upon detecting some specific errors only. We can interpret that in non-attack static network scenario less DAOs will be transmitted due to stable network, and therefore DAO insider attack can be easily detected in static networks. But in case of mobile network where network is very dynamic and susceptible to frequent changes causing more number of DAOs to be transmitted in non-attack situation, therefore it is difficult to distinguish between non-attack and under-attack condition in mobile networks. We have shown that the Li-MSD is able to reduce false positives as compared to SecRPL (as discussed in section \ref{analysis of fpr}. Li-MSD is enabled with blacklist table which helps in achieving early mitigation of attack after first instance of attack is detected. Li-MSD is capable of performing attack detection and mitigation in mobile environment also (as discussed in sections \ref{PDR-1}, \ref{PDR-2}, \ref{PDR-3}, \ref{PDR-4}).)   } 

\begin{table*}[!h]
	\centering
	\caption{Symbols and Definitions}
	\label{Symbols}
	\begin{tabular}{|p{6cm}|p{6cm}|}
		\hline
		\textbf{Symbol} & \textbf{Definition}\\ \hline
		
		\textit{$Node_{max}$} & Maximum number of nodes in the network. \\ \hline
		\textit{N$_{blacklist}$} & Number of Blacklisted nodes. \\ \hline
		\textit{T$_{child}$} & Number of nodes in neighbor table. \\ \hline
		\textit{$\mathcal{Z}$ $ \gets $ $ \textit{[}1,\dots,Node_{max}\textit{]} $} & Blacklist table \\ \hline
		\textit{$\mathcal{Q}$ $ \gets $ $ \textit{[}1,\dots,Node_{max}\textit{]}  $ } & Neighbor table \\ \hline
		\begin{tabular}[c]{@{}c@{}}\\ \textit{$\aleph_i$ $ \gets $ \textit{[}$ < $$ bl_{src_{ip}}$$>$\textit{]},} \\{$i  = $ $ 1 $,\ldots, $Node_{max}$}\end{tabular} & Structure of a blacklisted node entry in blacklist table. Where, $ bl_{src_{ip}}$ represents the blacklisted node IP address \\ \hline
		\begin{tabular}[c]{@{}c@{}}\\ \textit{$\Upsilon_i$ $\gets$ [$<$from, $global_{id}$, $DAO_{count}$$>$],} \\
{$i = 1,\ldots,Node_{max}$}\end{tabular} & Structure of a node entry in neighbor table. Where, \textit{from} represents the child node, $global_{id}$ represents the Global Prefix of child, $ DAO_{count} $ represents the total number of DAO's received from that child till current time. \\ \hline
		\textit{sender$_{ip} $} & Source IP address of DAO sender node. \\ \hline
		\textit{$dao_{prefix}$} & DAO Prefix of DAO sender node.\\
		\hline
		\textit{$\beta$} & DAO receive threshold \\ \hline
	\end{tabular} 
\end{table*}

\subsection{Description of Li-MSD algorithm}
Pseudocode of Li-MSD procedure is presented in Algorithm \ref{Mitigation procedure}. Li-MSD detection logic is incorporated in the DAO processing method of ContikiRPL, which is executed every time the DAO message is received from any child node. DAO processing method processes the incoming DAO messages and executes corresponding routing operations like storing, updating, and forwarding depending on the MOP of the RPL instance. Li-MSD attack mitigation procedure starts with the execution of INITIALIZE procedure. INITIALIZE procedure is responsible for initializing the required data structures and variables. Note that INITIALIZE procedure is called only once in 1800 seconds. A detailed description of INITIALIZE procedure is done in Section \ref{Desc:init}. Then, the parent node checks whether the DAO sender's address(\textit{ sender$_{ip} $}) is already present in Blacklist table (\textit{$\mathcal{Z}$}) or not. If \textit{ sender$_{ip} $} matches with any blacklisted node's address, this means that parent had already detected that DAO sender as an attacker node earlier, and it simply discards the received DAO message without any further processing. This not only saves the energy of nodes but also helps in the quick mitigation of attacks. In case \textit{$sender_{ip}$} is not present in \textit{$\mathcal{Z}$}, then the mitigation logic starts checking the Neighbor table (\textit{$\mathcal{Q}$}) to find out \textit{ sender$_{ip} $}. If \textit{ sender$_{ip} $} is not present in \textit{$\mathcal{Q}$}, it implies that the DAO sender is a new child node that has sent a DAO message for the first time. A new node entry in \textit{$\mathcal{Q}$} is created, and the DAO sender's information is added to \textit{$\mathcal{Q}$}. Corresponding to every child node entry in \textit{$\mathcal{Q}$} three values are stored, i.e., $from, global_{id}$, and $DAO_{count}$. Based on these entries, the mitigation logic decides whether a DAO sender is an attacker or not. It is important to note that whenever a node generates a DAO message, it appends its global ID (i.e., global IPv6 address) in that message. In RPL terminology, the DAO sender's global ID ($global_{id}$) is represented as the DAO prefix ($dao_{prefix}$).

\subsubsection{\textcolor{black}{Solution for Victim Node Problem}}
In Li-MSD, the DAO prefix (a variable of RPL protocol) is used to increment the DAO counter value (i.e., $DAO_{count}$) corresponding to that child and prevent the ``The Victim Node Problem". Whenever any parent node receives a DAO message from its child node, there are two ways in which DAO is handled. In the first case, if the child node is the DAO originator (i.e., $dao_{prefix}$ equals $global_{id}$), then $DAO_{count}$ corresponding to that child node is incremented, and the DAO message is forwarded to the parent. In the second case, when the child is not the DAO originator (i.e., $dao_{prefix}$ does not equal $global_{id}$), the corresponding $DAO_{count}$ is not incremented, and DAO message is forwarded to the parent. With this methodology, the mitigation logic detects attacker nodes correctly and avoids blacklisting of victim nodes. If any DAO originator is transmitting DAO messages very frequently, then the parent of that node will increment the $DAO_{count}$. When the value of $DAO_{count}$ corresponding to any child becomes equal to $\beta$, then the parent of that node directly blocks it and adds its information in the \textit{$\mathcal{Z}$}. \textcolor{black}{$dao_{prefix}$ and $global_{id}$ are RPL protocol variables which are utilized to solve ``The Victim Node Problem".  }


\subsection{Description of INITIALIZE procedure}\label{Desc:init}
\textcolor{black}{Pseudocode of INITIALIZE procedure is depicted in Algorithm \ref{Initialization procedure}}. This procedure is called when the Li-MSD procedure is executed for the first time after the node is turned ON, then it is then after every 1800 seconds interval. Re-initialization helps free up allocated memory and reduces false positives. in  The main responsibility of INITIALIZE procedure is to initialize the variables ($Node_{max}$, $\beta$, \textit{N$_{blacklist}$}, \textit{T$_{child}$}) and data structures ($\mathcal{Z}$, $\mathcal{Q}$, $\aleph$, $\Upsilon$) involved in Li-MSD algorithm. 

\subsection{Description of SEARCH-BLACKLIST procedure}\label{Desc:black}
SEARCH-BLACKLIST procedure \textcolor{black}{(Algorithm \ref{Search blackilist procedure})} plays an important in improving the performance of Li-MSD. This procedure helps in the early mitigation of attacks by detecting malicious DAO messages to avoid unnecessary processing of packets received from the attacker node. \textcolor{black}{SEARCH-BLACKLIST procedure is called when any node receives DAO from its child node. DAO $sender_{ip}$ is searched in $\mathcal{Z}$, and if found, then the DAO message is discarded without any further processing which achieves quick blocking of malicious packets. This procedure play an important role in achieving the early mitigation.}

\begin{algorithm}[!h]
	\begin{algorithmic}[1]
		\caption{Pseudocode of Li-MSD}
		\label{Mitigation procedure}
		\State call INITIALIZE procedure
		\Procedure{Li-MSD()}{}
		\If {SEARCH-BLACKLIST($sender_{ip}$) = TRUE} 
		\State discard the received DAO
		\State \textbf{return} \Comment{In case the sender node was already blacklisted}
		\EndIf
		\For {\textit{i $ \gets $ $ 1 $, T$ _{child} $}} 
		\If{\textit{$\mathcal{Q}\textit{[}\Upsilon_i.from\textit{]}$ $ = $ sender$_{ip} $}}
		\If {$sender_{ip}.dao_{prefix}$ $=$ $\mathcal{Q}\textit{[}\Upsilon_i.global_{id}\textit{]}$}
		\If {$\mathcal{Q}\textit{[}\Upsilon_i.DAO_{count}\textit{]}$ $<$ $\beta$}
		\State $\mathcal{Q}\textit{[}\Upsilon_i.DAO_{count}\textit{]}++$
		\State Forward DAO to preferred parent
		\Else
		\State Add $sender_{ip}$ in $\mathcal{Z}$
		\EndIf
		\Else
		\State Forward DAO to preferred parent
		\EndIf
		\Else
		\State Add and initialize new neighbor's information in $\mathcal{Q}$
		\EndIf      
		\EndFor
		
		\EndProcedure
	\end{algorithmic}
\end{algorithm}

\begin{algorithm}[!h]
	\begin{algorithmic}[1]
		\caption{Pseudocode of initialization procedure}
		\label{Initialization procedure}
		\Procedure{Initialize()}{}
		\State $Node_{max}$, $\beta$
		\State $\mathcal{Z}$ $ \gets $ $ \textit{[}1,\dots,Node_{max}\textit{]} $, $\mathcal{Q}$ $ \gets $ $ \textit{[}1,\dots,Node_{max}\textit{]}  $
		\State \textit{$\aleph_i$ $ \gets $ \textit{[}$ < $$ bl_{src_{ip}}$$>$\textit{]},} {$i  = $ $ 1 $,\ldots, $Node_{max}$}
	\State $\Upsilon_i \gets [<\textit{from}, \textit{global}{id}, \textit{DAO}{count}>],$ \hspace{10pt} $i = 1, \ldots, \textit{Node}_{\textit{max}}$
		\EndProcedure
	\end{algorithmic}
\end{algorithm}

\begin{algorithm}[!h]
	\begin{algorithmic}[1]
		\caption{Pseudocode of blacklist search procedure}
		\label{Search blackilist procedure}
		\Procedure{SEARCH-BLACKLIST($sender_{ip}$)}{}
		\For{\textit{i $ \gets $ $ 1 $, N$_{blacklist}$}}
		\If{$\mathcal{Z}\textit{[}\aleph_i.bl_{src_{ip}}\textit{]}$ $ = $ $sender_{ip} $} \Comment Early mitigation
		\State \Return TRUE
		\Else
		\State \Return FALSE
		\EndIf
		\EndFor
		\EndProcedure
	\end{algorithmic}
\end{algorithm}

\subsection{Time Complexity of Li-MSD}

\begin{itemize}
	\item \textcolor{black}{The lines $2-5$ of INITIALIZE procedure takes constant time, i.e., $O(1)$ since they only initialize the variables, structures, thresholds, neighbor table, and blacklist table, correspondingly.}
	
	\item In the SEARCH-BLACKLIST procedure, the blacklist table is investigated to check if any unauthorized senders have already been blacklisted or not. \textcolor{black}{If the size of blacklist table is $N_b$ and unauthorized senders are present in the blacklist table, then lines $2-8$ will take at most $O(N_b)$ time to find unauthorized senders.}
	
	\item \textcolor{black}{Li-MSD procedure examines the neighbor table to find unauthorized senders and identify malicious senders. After identifying the unauthorized senders, they are added to the blacklist table. Line $1$ calls INITIALIZE procedure which takes at most $O(1)$ time. Line $3$ finds the $sender_{ip}$ in blacklist table by calling SEARCH-BLACKLIST procedure which takes at most $O(N_b)$ time. Line $4$ executes in $O(1)$. Suppose the size of the neighbor table is $T_n$. Lines $7-22$ are responsible for finding and adding an unauthorized sender to the blacklist table and take $O(N_b) + O(T_n)$ in worst time because entire neighbor table is explored after examining the complete blacklist table.}
\end{itemize}

The time complexity of the proposed Li-MSD will be $O(N_b)+O(T_n)+O(1)$, i.e., $O(N_b)+O(T_n)$ because time needed for the initialization is $O(1)$. 

\section{Performance Evaluation}\label{Performance Evaluation}

\subsection{Experimental Setup}

Li-MSD is implemented by modifying the core files of ContikiRPL \cite{dunkels2004contiki}. The experiments for performance evaluation of Li-MSD are done in Cooja Simulator \cite{osterlind2006cross}. Cooja includes an inbuilt hardware simulator named MSPsim that emulates the exact binary code of real micro-controllers to achieve realistic simulation results. In this paper, Zolertia 1 (Z1), i.e., msp430 based micro-controller, is used for running Contiki operating system \cite{advancare2010zolertia}. \textcolor{black}{The simulation parameters mentioned in Table \ref{tab:SIMtable} are considered for carrying out experiments. We have considered static as well as mobile network scenarios and analyzed the network performance by varying the rate of attack ($1, 2, 4, 8$ seconds). A random topology consisting of 15 legitimate client, 1 server, and 4 attacker nodes is considered with UDGM radio model. Random Waypoint Mobility Model is used to simulate mobile nodes which move at a speed of 1-2 m/s \cite{kabilan2018performance}. To mount the DAO insider attack, an attacker can compromise any legitimate node and reprogram it to capture the DAO message and then transmit the captured DAO message in a fixed time of interval. To avoid any biased results, in this paper the legitimate nodes have been chosen randomly for creating attacker node. Contiki is a open source operating system and can be easily manipulated to create attacker node firmware. For this purpose ``\textit{rpl-dag.c, rpl-timers.c, and rpl-icmp6.c}" files of Contiki are modified to create attacker with varying replay intervals. The DAO attack is launched after receiving a DIO message from any parent node. To simulate a real world scenario, the attacker node is programmed to launch an attack after 90 seconds of network initialization. In this way, the attack starts after the network is established and becomes stable. Similarly, Li-MSD is programmed to activate after 120 seconds of network initialization. Other experiment related configuration settings are mentioned in Table \ref{tab:SIMtable}.} The detection approach of the proposed solution is activated upon network initialization.  

\begin{table*}[!h]
	\centering
	\caption{Simulation Parameters}
	\begin{tabular}{ |p{5cm}|p{7cm} | }
		\hline
		Parameter & Values\\
		\hline
		Radio model & Unit Disk Graph Radio Medium (UDGM)\\
		\hline
		Mobility model & Random Waypoint Mobility Model\\
		\hline
		Network area & 150m × 150m\\
		\hline
		Simulation time  & 1800 s\\\hline
		Objective function(OF) & Minimum Rank with Hysteresis Objective Function(MRHOF)\\
		\hline
		Number of attacker nodes & 4\\
		\hline
		Number of server nodes& 1\\
		\hline
		Number of client nodes& 15\\
		\hline
		DAO Replay interval& 1, 2, 4, 8 seconds\\
		\hline
		Data packet size & 30 bytes\\
		\hline
		Data sending interval& 60 seconds\\
		\hline
		Transmission power& 0dBm\\
		\hline
		Mobile node speed & 1-2 m/s\\
		\hline
		
	\end{tabular}
	\label{tab:SIMtable}
\end{table*}

For all the scenarios, 10 independent replications with different random seeds were executed to obtain statistically correct experimental results. We have reported the average values of the collected results with their errors at a 95\% confidence interval to avoid any biased observations.

\subsection{Performance Metrics}
\textcolor{black}{To analyze the attack's impact on RPL and evaluate the performance of Li-MSD in static and mobile network scenarios, Packet Delivery Ratio (PDR) and Average End-to-End Delay(AE2ED), Average Power Consumption (APC), Packet Loss Ratio (PLR), False Positive Rate (FPR) are chosen. Memory Overhead of Li-MSD on Z1 mote is also studied. PDR, AE2ED, APC, and PLR are the most prominent metrics which are used in the literature for analyzing network performance. In this paper, the aim is to leverage network performance during the attack therefore study of PDR, AE2ED, APC, and PLR becomes necessary to show that Li-MSD is capable of improving network performance while the network is under DAO insider attack. Similarly, FPR shows the effectiveness of Li-MSD in terms of its capability to predict non-attack instances correctly. FPR analysis is important because high FPR results in legitimate nodes getting blocked, which consequently impacts network performance. Memory Overhead of Li-MSD is studied to analyze the memory requirements and check whether the proposed solution is suitable for resource-constrained nodes or not.}

These performance metrics are defined as,
\noindent

\begin{enumerate}
\color{black}
	\item \textbf{PDR: } It represents the ratio between the number of data packets successfully received at the server (DODAG root) to the number of data packets sent by client (sensor) nodes. 

	\item \textbf{AE2ED: } It represents the average time taken by a data packet to get successfully delivered at the server node. 
	\item \textbf{APC: } It is defined as the average of total power consumed by each non-root node in the given period of time. 
	\item \textbf{PLR:} It represents the ratio between the number of data packets not received at the server (DODAG root) to the number of data packets sent by client (sensor) nodes.
	\item \textbf{FPR: } False Positives are instances that are actually non-attack but are classified as attack. FPR is the ratio of actual non-attack instances that are incorrectly classified as attacks by the model, out of the total number of actual non-attack instances. 
	\item \textbf{Memory Overhead: } It represents the overhead of the proposed security solution induced on the node in terms of memory consumed. 

\end{enumerate}

\subsection{Simulation Results}
In this study, both static and mobile network scenarios have been considered for the performance evaluation of Li-MSD. The prominent performance metrics like PDR, AE2ED, APC, PLR, and Memory Overhead are utilized to show the effectiveness of our proposed solution. In a real-world scenario, the attacker can be aggressive or non-aggressive. Therefore we have reported results on different attack replay intervals, i.e., $ 1, 2, 4,$ and $8$ seconds. Performance of RPL (i..e., standard ContikiRPL implementation without attack), RPL\textsubscript{Under Attack} (i.e., RPL under \textcolor{black}{DAO insider attack}), RPL\textsubscript{Li-MSD} (i.e., RPL under attack with our proposed security solution), RPL\textsubscript{SecRPL} (i.e., RPL under attack with Ghaleb et al. \cite{ghaleb2018addressing} solution) is evaluated and compared. In the case of the RPL (reference implementation), it must be noted that the replay interval plays no role.

\subsubsection{Analysis of Packet Delivery Ratio (PDR)}\label{PDR-1}
\begin{figure}[!htb]
	\centering
		\centering
		\includegraphics[width=0.5\textwidth, trim={2cm 2cm 2cm 2cm}]{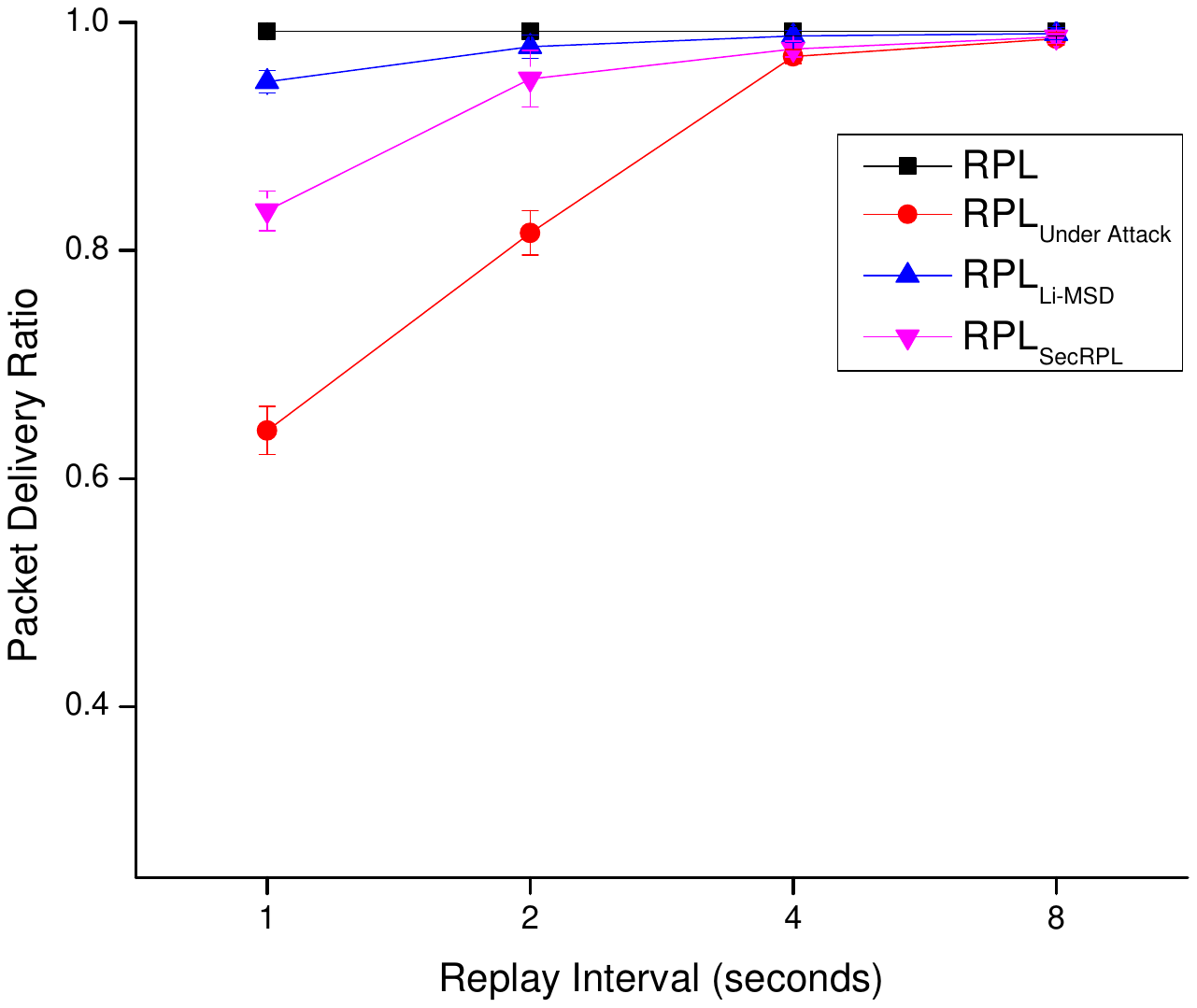}
		\caption{PDR values obtained in static scenario}
		\label{fig:PDR_static}
\end{figure}

\begin{figure}[!htb]
		\centering
		\includegraphics[width=0.5\textwidth, trim={2cm 2cm 2cm 2cm}]{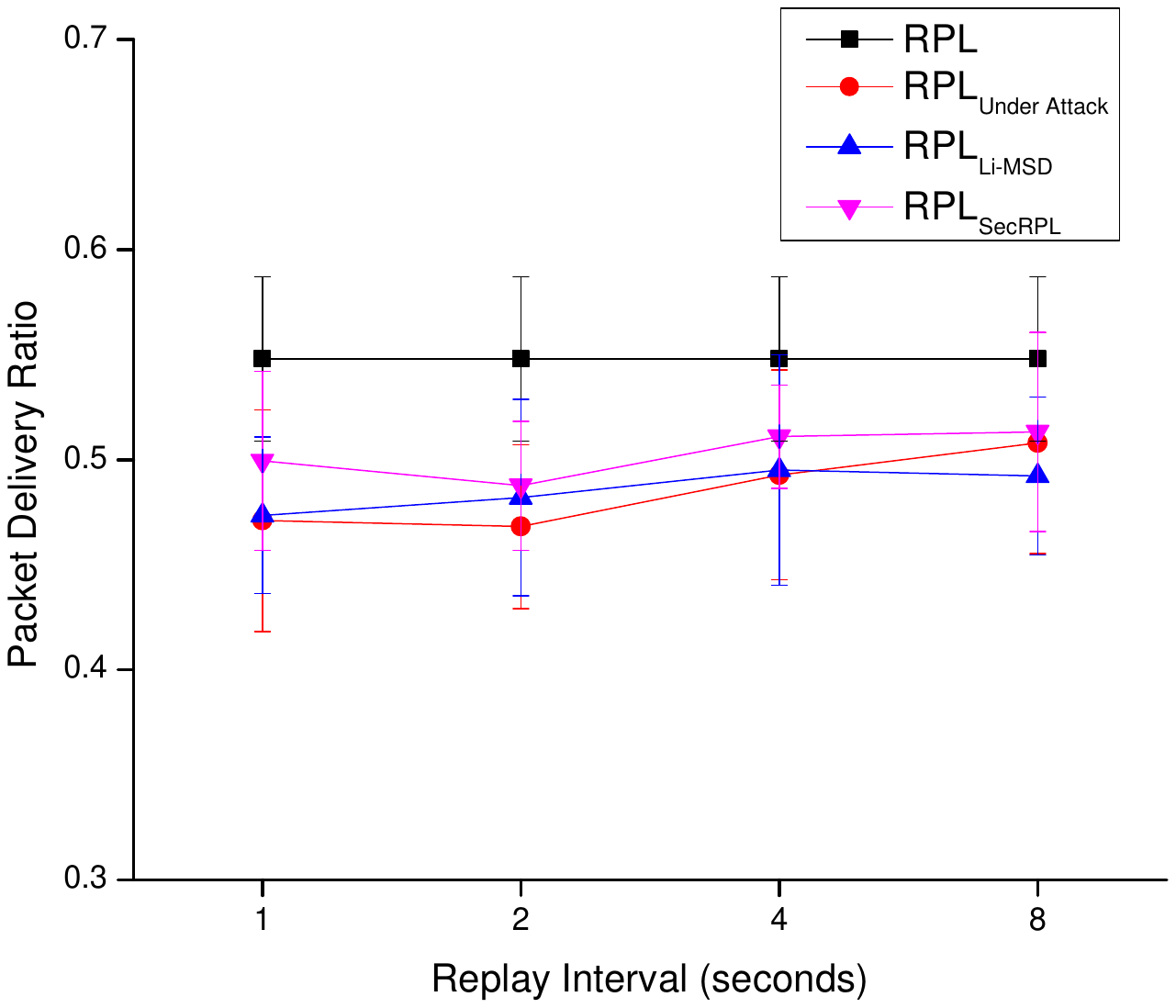}
		\caption{PDR values obtained in mobile scenario}
		\label{fig:PDR_mobile}
\end{figure}

PDR plays a vital role in safety-critical applications like healthcare and industrial monitoring. Therefore, it is very important to evaluate the effectiveness of Li-MSD in leveraging the network's PDR under attack conditions.
Fig. \ref{fig:PDR_static} and \ref{fig:PDR_mobile} shows PDR obtained with different replay intervals in case of static and mobile scenario, respectively.
It can be observed from the values of RPL\textsubscript{Under Attack} that DAO insider attack severely degrades the network PDR in both static and mobile scenarios. Fig. \ref{fig:PDR_static} indicates that Li-MSD is able to improve the performance of the network in the presence of aggressive ($1$ and $2$ seconds) as well as non-aggressive attackers ($3$ and $4$ seconds). Moreover, Li-MSD also outperforms SecRPL. The highest and lowest PDR achieved by RPL\textsubscript{Li-MSD} in a static scenario is $\approx$99\% and $\approx$96\%, respectively. This is because, in static scenarios, the physical locations of legitimate attackers do not change. That is why Li-MSD performs consistently. However, this is not the case in a mobile scenario, as indicated in Fig. \ref{fig:PDR_mobile}. When nodes move, they tend to have topology changes and frequent routing updates, drastically reducing PDR. From the experimental results, it is observed that Li-MSD can leverage network performance under attack conditions. Compared to the results of static scenarios, Li-MSD has lesser effectiveness in mobile scenarios. This is because a moving attacker is not blacklisted quickly by legitimate nodes, and in that time, the attacker node can affect a lot of nodes, which leads to reduced performance improvement. The highest and lowest PDR achieved by RPL\textsubscript{Li-MSD} in a mobile scenario is $\approx$48\% and $\approx$46\%, respectively. It is to be noted that in the mobile scenario, there are lot of variations in results reported by independent replications, which is indicated by errors bars. This is because of frequent changes in topology and network dynamics. It is concluded from Fig. \ref{fig:PDR_static} and \ref{fig:PDR_mobile} that Li-MSD effectively improves the network's performance as it can detect and mitigate DAO insider attacks in RPL-based IoT networks. 


\begin{figure}[!h]
	\centering
		\centering
		\includegraphics[width=0.5\textwidth, trim={2cm 2cm 2cm 2cm}]{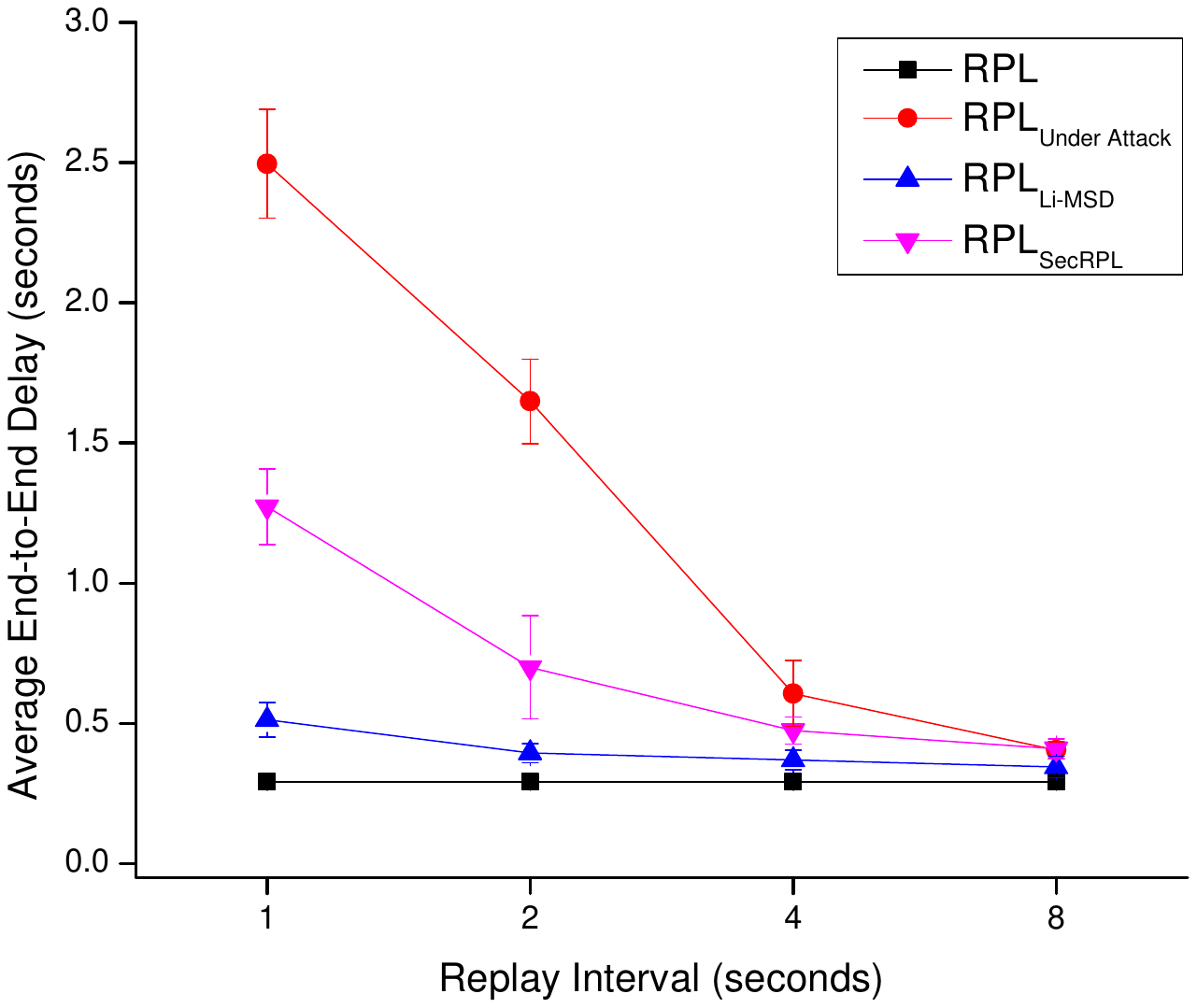}
		\caption{AE2ED values obtained in static scenario}
		\label{fig:AE2ED_static}
 \end{figure}
 \begin{figure}[!h]
		\centering
		\includegraphics[width=0.5\textwidth, trim={2cm 2cm 2cm 2cm}]{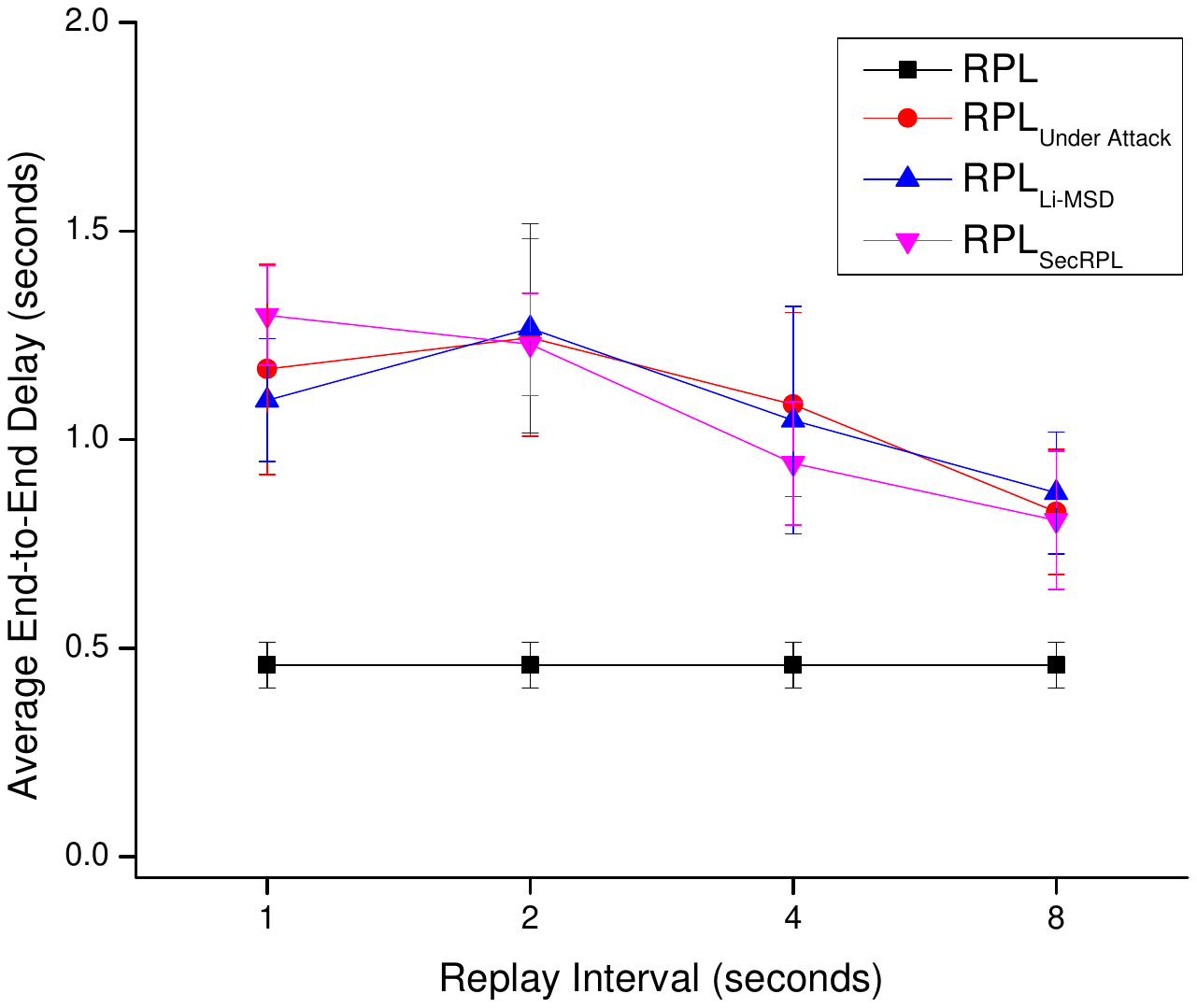}
		\caption{AE2ED values obtained in mobile scenario}
		\label{fig:AE2ED_mobile}
\end{figure}

\subsubsection{Analysis of Average End-to-End Delay (AE2ED)}\label{PDR-2}
Analysis of AE2ED is crucial because many IoT applications demand minimum packet delay. Therefore, it is important to develop a security solution capable of minimizing AE2ED in the network. Fig. \ref{fig:AE2ED_static} shows AE2ED values obtained in static scenario. It can be observed that RPL\textsubscript{Under Attack} that aggressive attacker has a severe impact on AE2ED as compared to a non-aggressive attacker because of the frequency of malicious DAO transmissions. Li-MSD shows significant improvement in values of AE2ED as compared to RPL\textsubscript{Under Attack}, and RPL\textsubscript{SecRPL}. The highest and lowest AE2ED achieved by RPL\textsubscript{Li-MSD} in a static scenario is $\approx$0.50 seconds and $\approx$0.30 seconds, respectively. The main reason for this significant performance improvement is that attacker nodes are blocked quickly after the launch of the attack. Therefore, legitimate nodes can process and forward child node's data packets in less time. This is valid for both aggressive as well as non-aggressive attack. Fig. \ref{fig:AE2ED_mobile} shows the results obtained in mobile scenario. \textcolor{black}{The highest and lowest AE2ED achieved by RPL\textsubscript{Li-MSD} in the mobile scenario are $\approx$1.28 seconds and $\approx$0.85 seconds, respectively. The proposed solution improves the network's AE2ED in different replay intervals. Network dynamics play an important role in the attacker's impact on the network and the performance of the proposed solution. The mobile nature of nodes gives an extra advantage to the adversary to increase its attack range. Moreover, it also makes it tough for defense solutions to achieve acceptable performance under attack conditions. The experimental results conclude that Li-MSD can improve the network's AE2ED in both static and mobile scenarios by blocking the attacker and reducing the attack's impact on legitimate nodes.}

\subsubsection{Analysis of Average Power Consumption (APC)}\label{PDR-3}
\begin{figure}[!h]
	\centering
		\centering
		\includegraphics[width=0.5\textwidth, trim={2cm 2cm 2cm 2cm}]{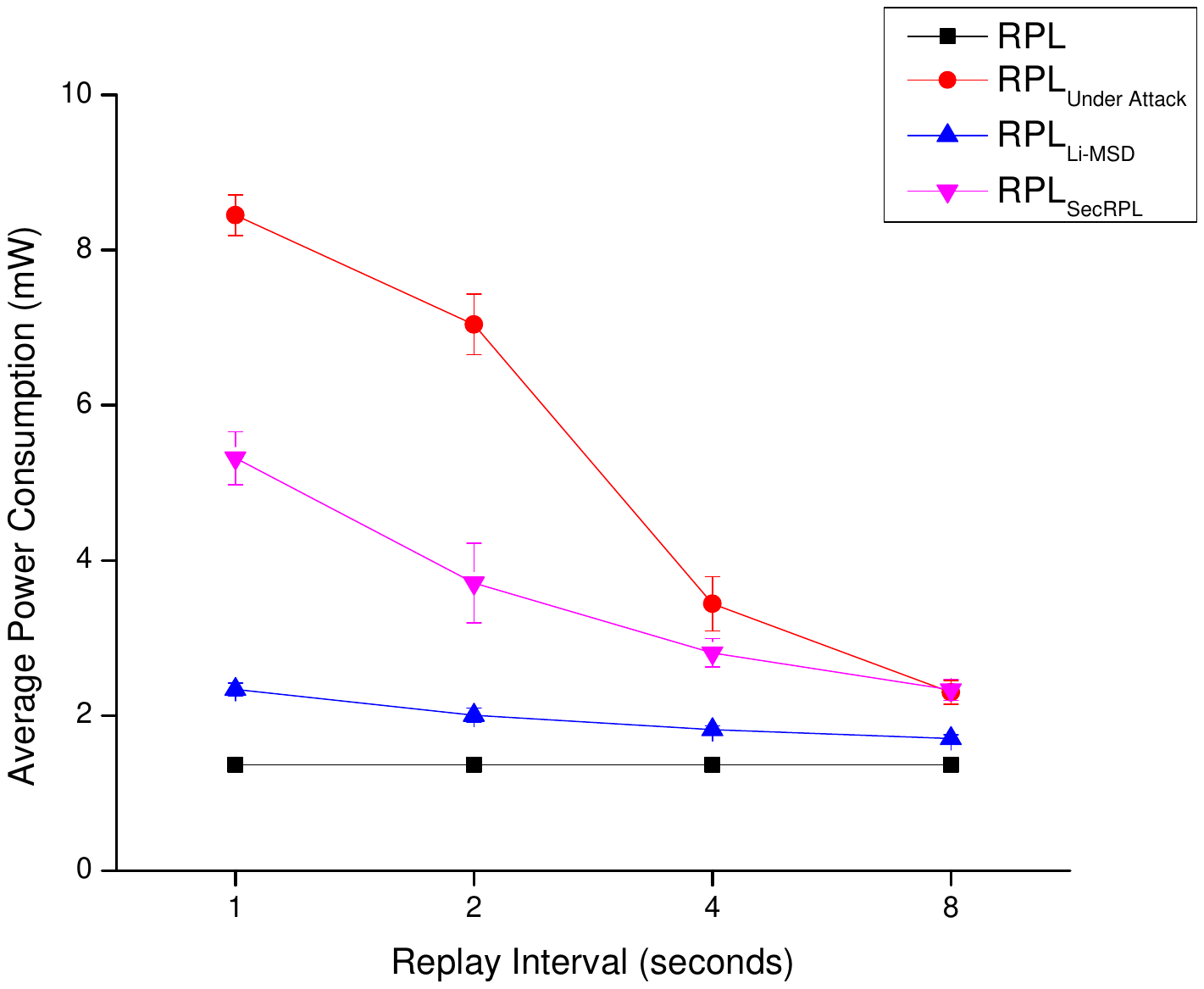}
		\caption{APC values obtained in static scenario}
		\label{fig:APC_static}
\end{figure}

\begin{figure}[!h] 
		\centering
		\includegraphics[width=0.5\textwidth, trim={2cm 2cm 2cm 2cm}]{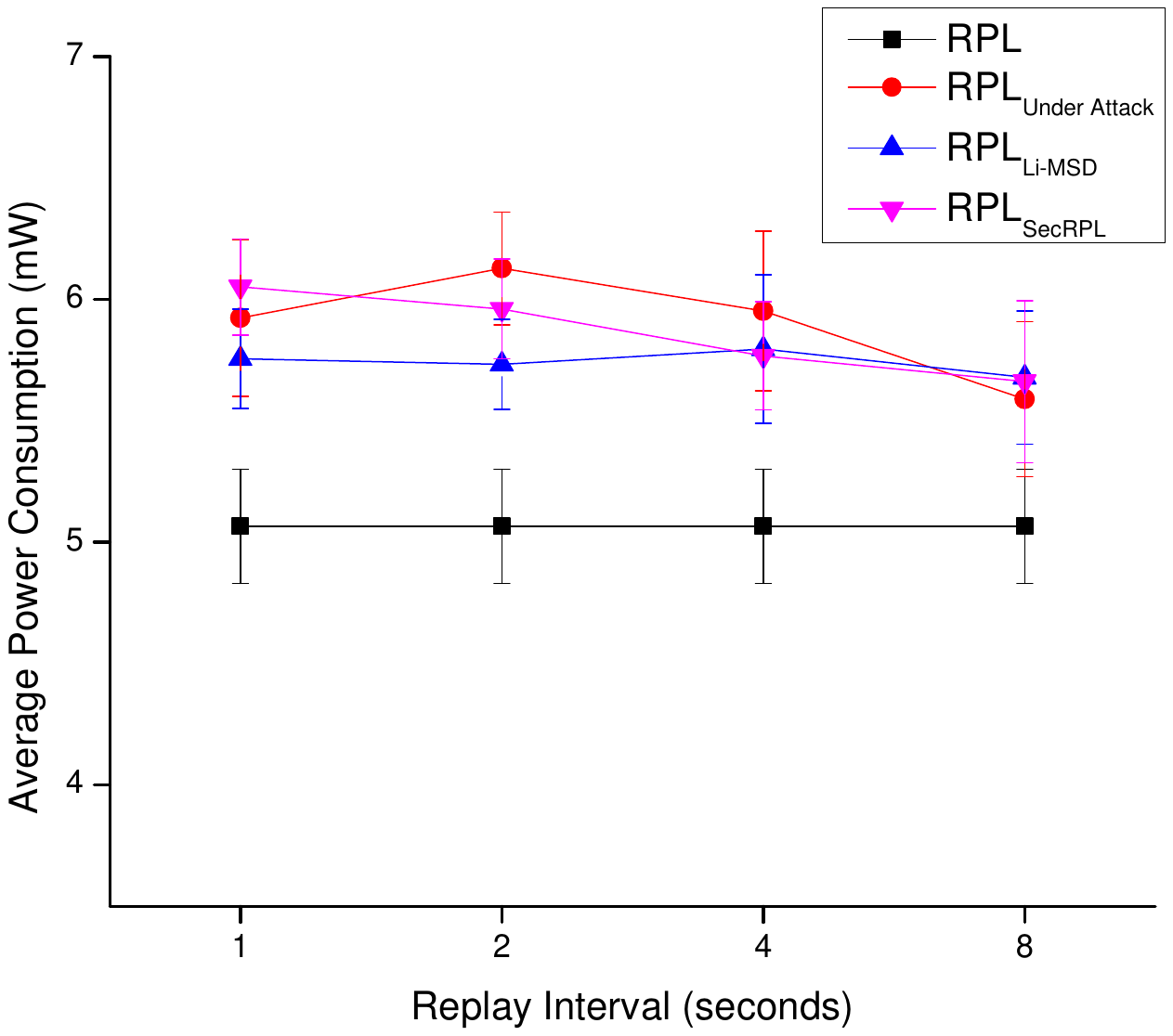}
		\caption{APC values obtained in mobile scenario}
		\label{fig:APC_mobile}
\end{figure}

RPL is widely used because it provides energy-efficient routing in LLNs. Therefore, it is very important to study the power consumption of nodes when any new security solution is deployed on them. Fig. \ref{fig:APC_static}, \ref{fig:APC_mobile} present the APC values obtained in case of static and mobile network scenario, respectively. The values indicated by RPL\textsubscript{Under Attack} clearly show the impact of attack on the network. APC is inversely proportional to network lifetime. Therefore, this metric needs special attention while designing a security solution. Li-MSD modules are capable of reducing the APC of network nodes in the presence of attackers. The values represented by RPL\textsubscript{Li-MSD} in Fig. \ref{fig:APC_static} show that the proposed solution can improve the network's performance in static scenarios significantly. Moreover, the results obtained in mobile scenarios also conclude the same. Li-MSD outperforms SecRPL in both static and mobile network scenarios in the presence of aggressive as well as non-aggressive attackers. The improved values of APC indicate that Li-MSD effectively leverages network lifetime.


\begin{figure}[!h]
	\centering
		\centering
		\includegraphics[width=0.5\textwidth, trim={2cm 2cm 2cm 2cm}]{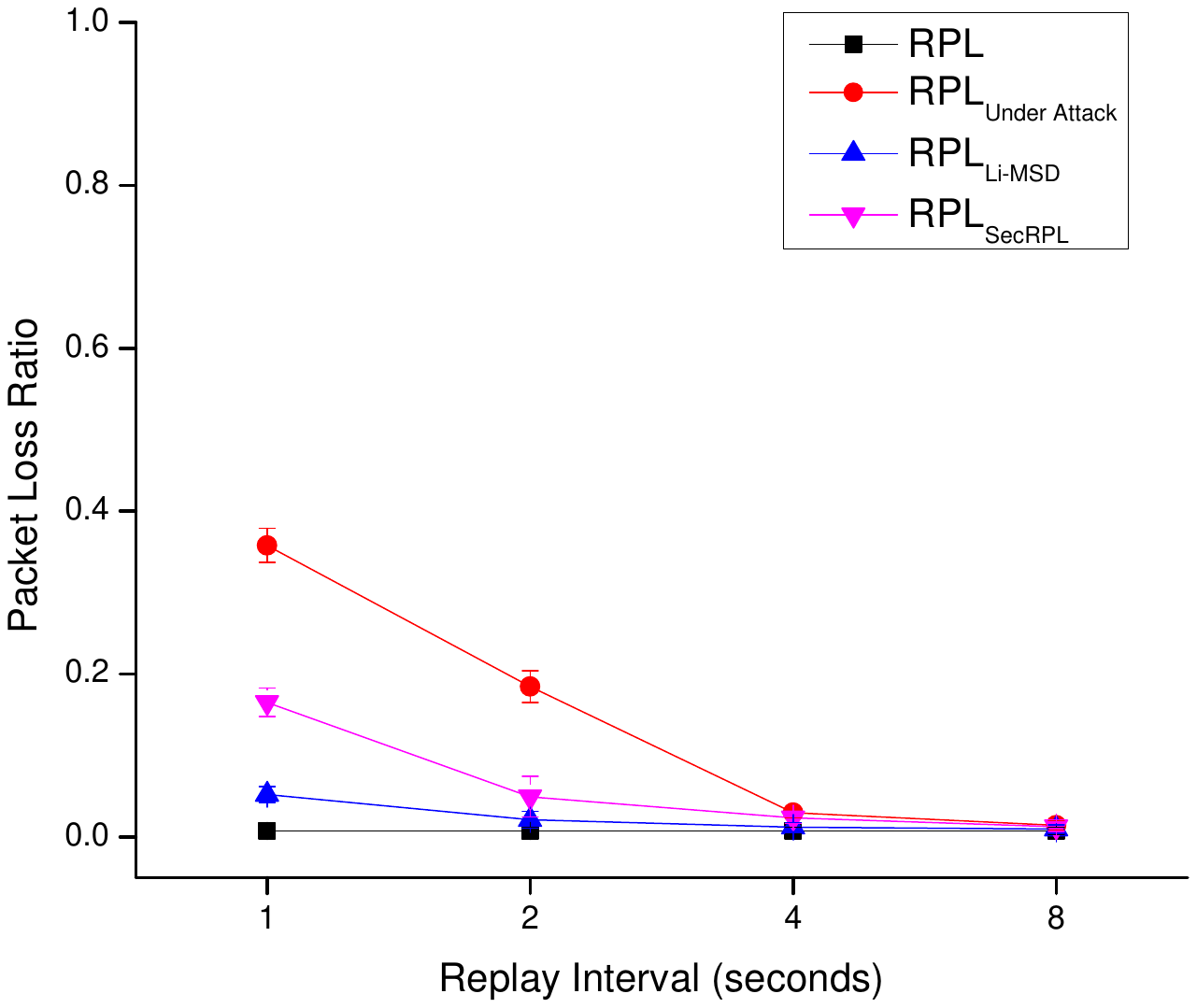}
		\caption{PLR values obtained in static scenario}
		\label{fig:PLR_static}
\end{figure} 
\begin{figure}
		\centering
		\includegraphics[width=0.5\textwidth, trim={2cm 2cm 2cm 2cm}]{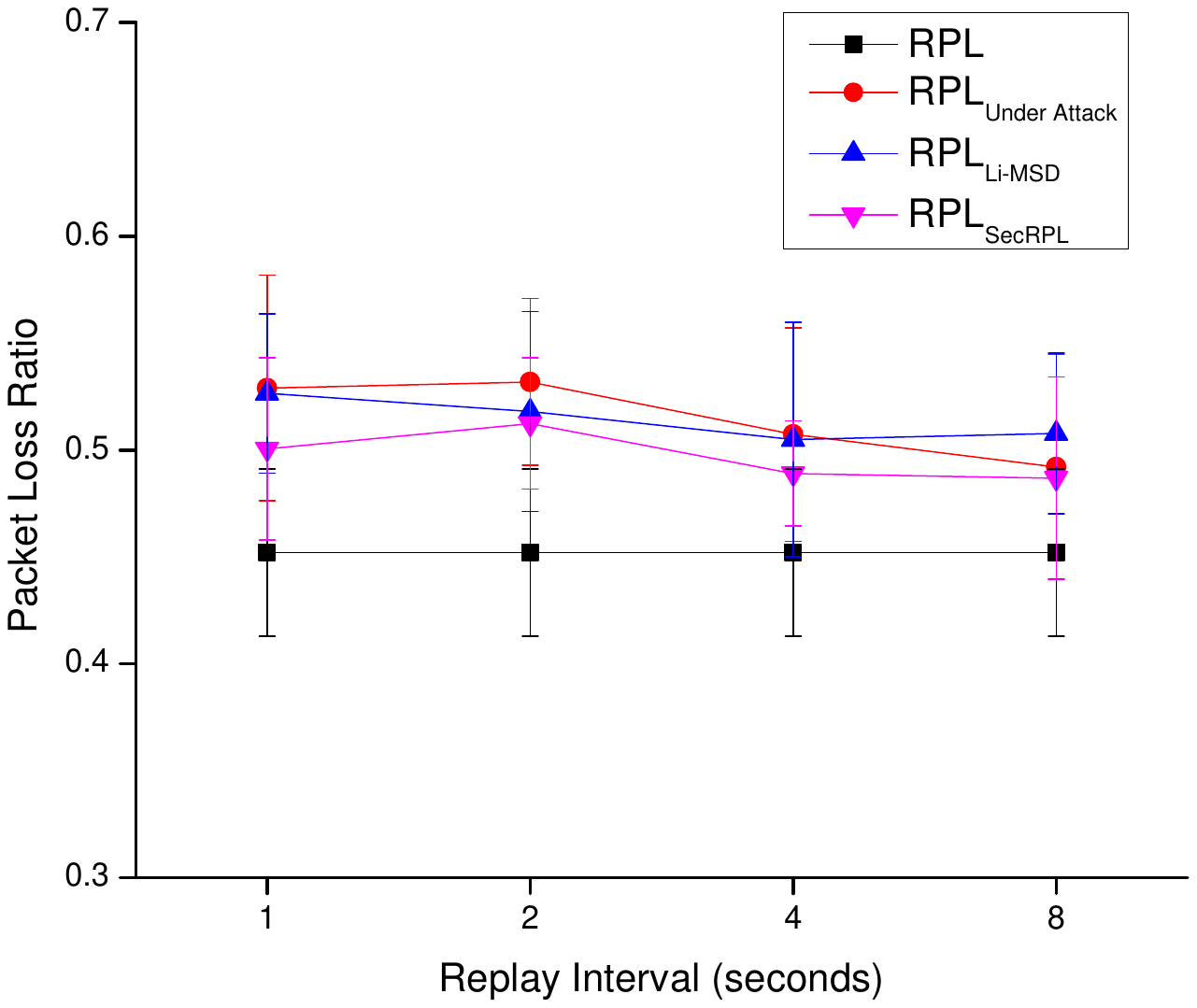}
		\caption{PLR values obtained in mobile scenario}
		\label{fig:PLR_mobile}
\end{figure}

\subsubsection{Analysis of Packet Loss Ratio (PLR)}\label{PDR-4}
In this paper, packet losses are studied in terms of PLR. Fig. \ref{fig:PLR_static} shows the PLR obtained in static scenario. In case of an aggressive attack the PLR is highly affected, which can be easily seen from the values of RPL\textsubscript{Under Attack} at 1 and 2 seconds replay interval. The non-aggressive attack (i.e., 3 and 4 seconds replay interval) has less impact on the network's PLR. From the results indicated in Fig. \textsubscript{Li-MSD} it can be observed that Li-MSD improves the PLR significantly. RPL\textsubscript{Li-MSD} achieves PLR values almost equal to RPL. Fig. \ref{fig:PLR_mobile} depicts the PLR values obtained in the mobile network scenario. RPL\textsubscript{Under Attack} values show the impact of DAO insider attack on RPL. Packet losses are high in the mobile scenario because of node mobility. Further, this loss increases in the presence of a DAO insider attacker. The PLR values of RPL\textsubscript{Under Attack} indicate that attackers can reduce network performance. Li-MSD can reduce the attack impact, as shown by mean and error bars. In mobile scenarios, by observing the values indicated by error bars, we conclude that Li-MSD achieves acceptable performance.   

\subsubsection{Analysis of Memory Overhead}

\begin{figure}[!h]
	\centering
	\includegraphics[width=.5\textwidth, trim={2cm 2cm 2cm 2cm}]{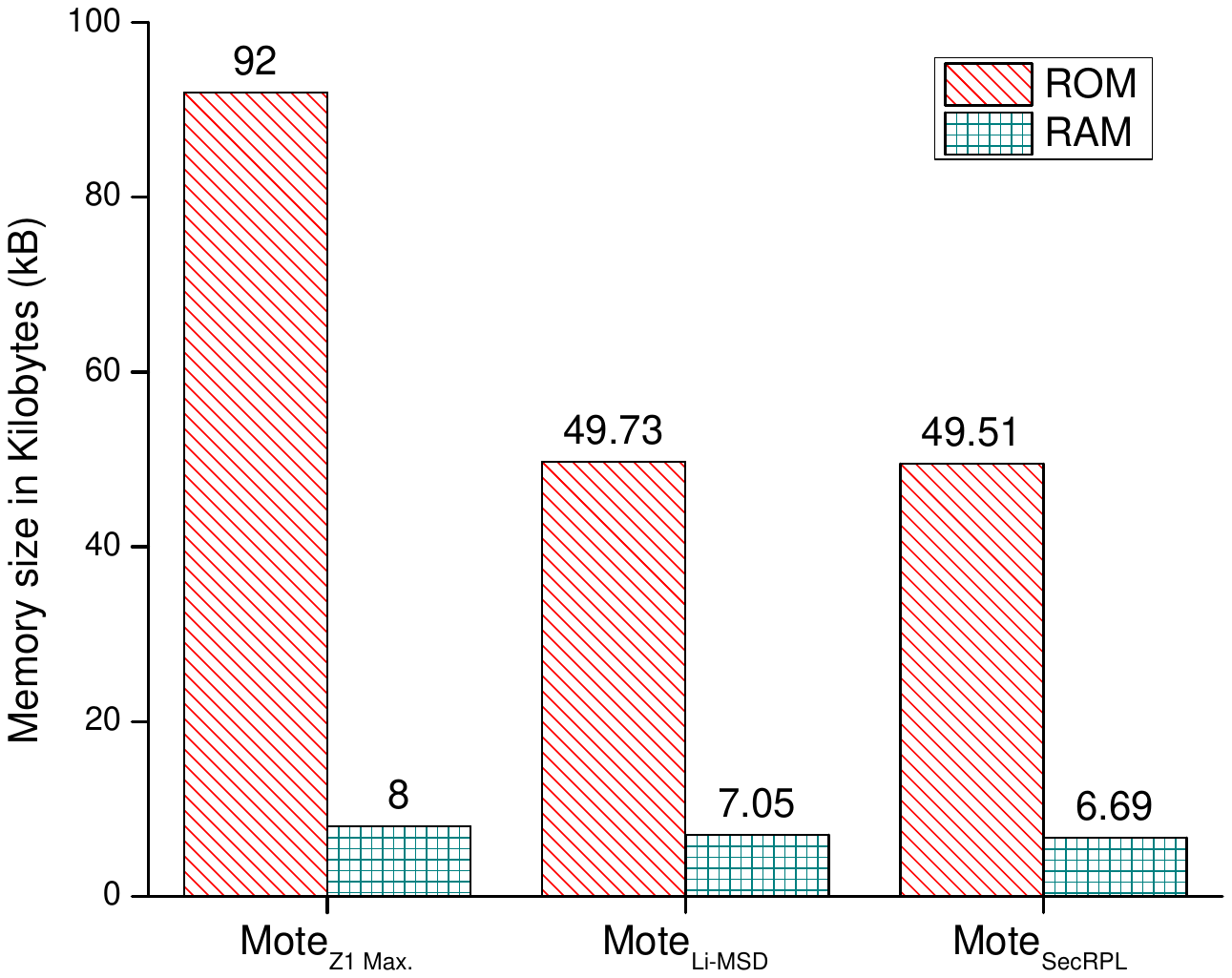}
	\caption{Memory Overhead}
	\label{fig:Memory_overhead}
\end{figure}

In LLNs, the usage of resource-consuming security solutions is not recommended. Therefore lightweight security solutions are developed to leverage the performance of such networks. Lightweight solutions do not impose significant overhead on resource-constrained nodes. This paper analyzes the implementation overhead of Li-MSD in terms of memory requirements (i.e., RAM and ROM) using the msp430-size tool. Fig. \ref{fig:Memory_overhead} shows comparison of RAM and ROM requirement of $Mote_{Li-MSD}$ (Contiki firmware with Li-MSD implemented), $Mote_{SecRPL}$ (Contiki firmware with SecRPL implemented) with $Mote_{Z1 Max.}$ (Maximum capacity of standard Z1 mote). Fig. \ref{fig:Memory_overhead} indicates that our proposed solution easily fits into Z1 motes without imposing significant overhead. It can be concluded that Li-MSD is a lightweight solution and suitable for resources-constrained networks. 

\subsubsection{Analysis of FPR}\label{analysis of fpr}

\begin{figure}[!h]
	\centering
	\includegraphics[width=.5\textwidth, trim={2cm 2cm 2cm 2cm}]{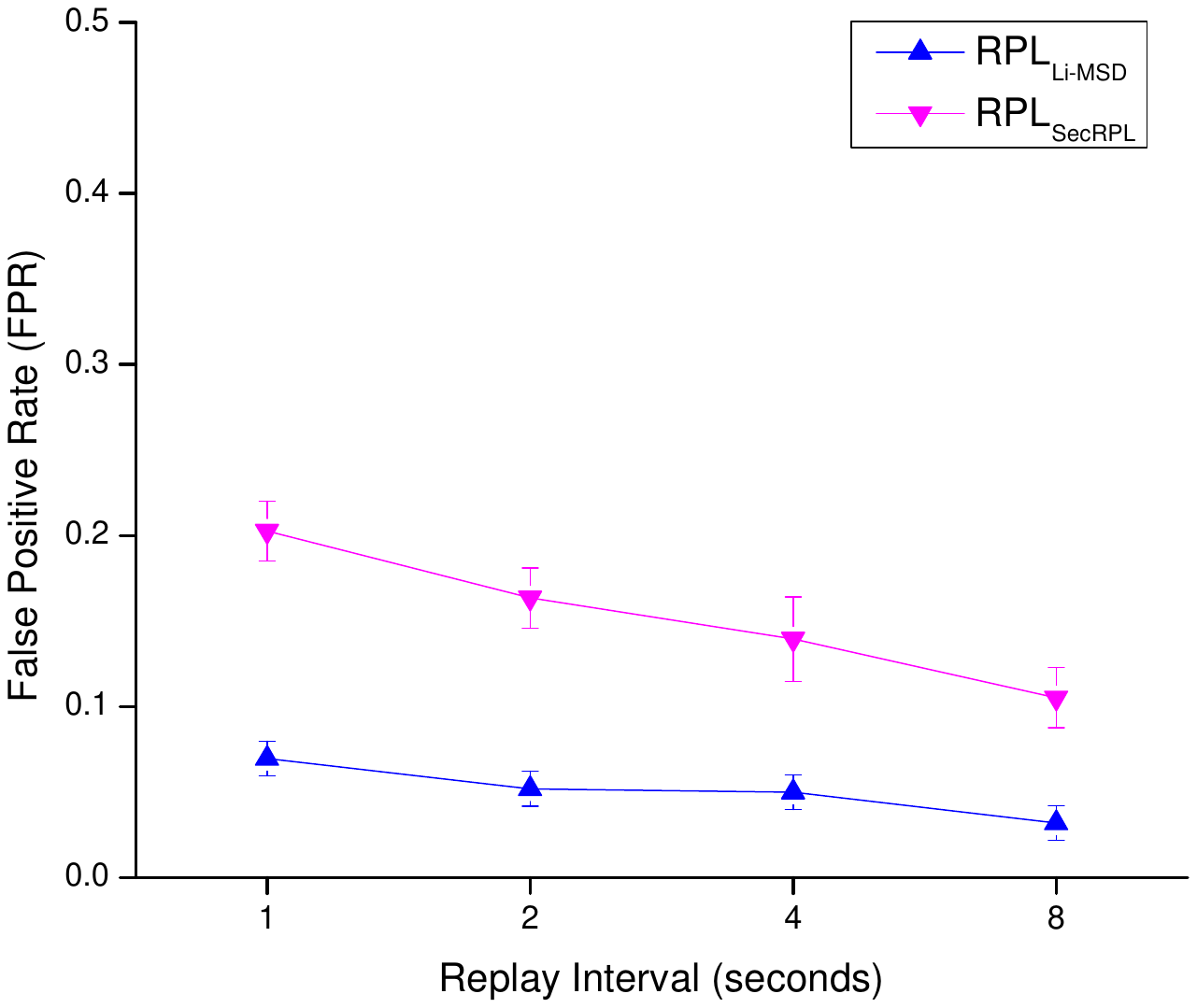}
	\caption{Comparison of FPR}
	\label{fig:FPR}
\end{figure}

\textcolor{black}{Figure \ref{fig:FPR} shows a comparison of the FPR obtained in the case of RPL\textsubscript{Li-MSD} and RPL\textsubscript{SecRPL}. The results indicated in Figure \ref{fig:FPR} demonstrate that RPL\textsubscript{Li-MSD} significantly improves the FPR compared to RPL\textsubscript{SecRPL}. The high FPR values of RPL\textsubscript{SecRPL} in all the replay intervals indicate that non-attack events are also classified as attack events, leading to the blocking of legitimate nodes. This situation may impact the performance of the networks; therefore, the FPR needs to be as minimal as possible. RPL\textsubscript{Li-MSD} is able to reduce the FPR and prevent the blocking of legitimate nodes.  }

\subsubsection{\textcolor{black}{Limitations of Li-MSD}}

\begin{enumerate}
	\color{black}
	\item \textbf{Reduced performance in mobile scenarios: } Li-MSD performance better in static network scenarios as compared to mobile scenarios. This is because, in static scenarios, the physical locations of legitimate attackers remain same therefore Li-MSD performs consistently. However, when nodes are mobile, they tend to have topology changes and frequent routing updates which drastically affects network performance. Moreover, a moving attacker is not blacklisted quickly by legitimate nodes, and in that time, the attacker node can affect a lot of nodes, which leads to reduced performance improvement in mobile scenarios.    
	\item \textbf{Static threshold: } Li-MSD uses static threshold for setting the safety parameter (DAO count ($\beta$)) which must be appropriately configured to achieve good performance of Li-MSD.   
	\item \textbf{Searching time of neighbor and blacklist tables:} Due to use of simple data structure (i.e., arrays) the searching time of nodes id's in neighbor and blacklist tables is high. This affects Li-MSD algorithm attack detection time.   
\end{enumerate}

\section{Conclusion and Future Scope}\label{Conclusion and Future Scope}

\textcolor{black}{RPL is still in its development stage and has several security vulnerabilities that an attacker may exploit to compromise network security. Therefore, the security of RPL-based applications is one of the utmost important concerns. In this paper, DAO insider attack is studied.} Using simulation-based study, we have shown that DAO insider attack can negatively impact the network performance in terms of parameters like packet delivery ratio, average end-to-end delay, average power consumption, and packet loss ratio. To address DAO insider attacks, we have proposed a lightweight security solution named Li-MSD. The experimental results indicate that Li-MSD effectively detects and mitigates the attack in static and mobile network scenarios. Li-MSD outperforms the existing solution in most cases without imposing any significant overhead on resource-constrained nodes like Zolertia Z1. \textcolor{black}{In addition, Li-MSD outperforms current state-of-the-art (i.e., SecRPL) in terms of false positive rate and other network metrics (specifically in all static network scenarios, and in some mobile network scenarios). Li-MSD performs significantly better in static networks than in mobile networks, that is its main limitation. In the future, we aim to improve the performance of Li-MSD in mobile networks by incorporating dynamic thresholding mechanism, optimize searching time to reduce overall packet processing time, and extend the detection logic to mitigate other DAO-based attacks. Popular security methods like message authentication, community based trust models, and access control offer a roadmap for addressing DAO-based attacks. Considering the limitations of existing defense mechanisms in RPL the implications for future security mechanisms for IoT security include designing lightweight security solutions, performance optimization in mobile networks, continuous improvement and adaptation, adoption of a holistic and multifaceted approach.}

\printcredits

\section*{Declaration of competing interest}
The authors declare that they have no known competing financial interests or personal relationships that could have appeared to influence the work reported in this paper.

\bibliographystyle{ACM-Reference-Format}

\end{document}